\documentclass{aa}

\usepackage{natbib}
\usepackage{graphicx}
\usepackage{subfigure}
\def\Ha{H$\alpha$}
\def\Hb{H$\beta$}

\def\Hd{H$\delta$}
\def\Oii{{\sc [Oii]}}

\def\Oiii{{\sc [Oiii]}}
\def\wings{{\sc wings}}
\def\sdss{{\sc sdss}}
\begin{document}

\title{A spectrophotometric model applied to cluster galaxies: the {\sc wings} dataset}

\author{J. Fritz\inst{1,2} \and B.~M. Poggianti\inst{1} \and D. Bettoni\inst{1} \and A. Cava\inst{1} \and W.~J. Couch\inst{3} \and M. D'Onofrio\inst{2} \and A. Dressler\inst{4} \and G. Fasano\inst{1} \and P. Kj\ae rgaard\inst{5} \and M. Moles\inst{6} \and J. Varela\inst{1}.}
\offprints{Jacopo Fritz,\\ 
           \email{jacopo.fritz@gmail.com}}
\institute{INAF-Osservatorio Astronomico di Padova, vicolo Osservatorio 5, 35122 Padova, Italy\\
\email{jacopo.fritz@gmail.com}
\and
 Dipartimento di Astronomia, vicolo Osservatorio 2, 35122 Padova, Italy
\and
 School of Physics, University of New South Wales, Sydney 2052, Australia
\and
 Observatories of the Carnegie Institution of Washington, Pasadena, CA 91101, USA
\and
 Copenhagen University Observatory. The Niels Bohr Insitute for Astronomy Physics and Geophysics, Juliane Maries Vej 30, 2100 Copenhagen, Denmark
\and
 Instituto de Astrof\'\i sica de Andaluc\'\i a (C.S.I.C.) Apartado 3004, 18080 Granada, Spain}
 
\date{Received ...; accepted ...}

\titlerunning{SFH in cluster galaxies}
\authorrunning{Fritz J. et al.}

\abstract
{The WIde-field Nearby Galaxy-cluster Survey (\wings) is a project aiming at the study of the galaxy populations in clusters in the local universe ($0.04<z<0.07$) and the influence of environment on the physical properties of galaxies. This survey provides a high quality set of spectroscopic data for $\sim 6000$ galaxies in $48$ clusters. The study of such a large amount of objects requires automatic tools capable of extracting as much information as possible from the data.}
{In this paper we describe a stellar population synthesis technique that reproduces the observed optical spectra of galaxies performing an analysis based on spectral fitting of the stellar content, extinction and, when possible, metallicity.}
{A salient feature of this model is the possibility of treating dust extinction as a function of age, allowing younger stars to be more obscured than older ones. Our technique, for the first time, takes into account this feature in a spectral fitting code. A set of template spectra spanning a wide range of star formation histories is built, with features closely resembling those of typical spectra in our sample in terms of spectral resolution, noise and wavelength coverage. Our method of analyzing these spectra allows us to test the reliability and the uncertainties related to each physical parameter we are inferring. The well-known degeneracy problem, i.e. the non-uniqueness of the best fit solution (mass and extinction in different age bins), can be addressed by assigning adequate error bars to the recovered parameters. The values found in this way, together with their error bars, identify the region of parameter space which contains all the possible solutions for a given spectrum. A comparison test was also performed on a \wings \ subsample, containing objects in common with the Sloan Digital Sky Survey, yielding excellent agreement.}
{We find that the stellar content as a function of age is reliably recovered in four main age bins and that the uncertainties only mildly depend on the S/N ratio. The metallicity of the dominant stellar population is not always recoverable unambiguosly, depending on the Star Formation History pattern.}
{}


\maketitle


\section{INTRODUCTION}

Galaxy spectra are a fundamental datum, carrying a wealth of information including the redshift and, via the measurement of spectral lines and continuum emission, provide the best insight of the stellar content. In particular, spectral synthesis has by now become a standard technique to derive the salient properties of the stellar populations in galaxies. 
Nowadays, the new simple stellar population (SSP, hereafter) spectra at medium and high resolution, together with the availability of more and more details in both observed stellar atmospheres and stellar evolution models, makes it much easier to reproduce an observed galactic spectrum as a combination of simple stellar population spectra. This in turn enables one to derive fundamental quantities that characterize the build up of stellar mass and the characteristics of stellar electromagnetic emission in a galaxy (i.e. Star Formation Rates, extinction, age of stellar populations, stellar masses).

The paper by \cite{poggianti01}, which represents the foundations of the present work, exploited the power of spectral synthesis as a tool to investigate the properties of the stellar populations in a galaxy, taking into account the information carried by the whole optical spectrum. Many techniques have been recently developed using the information which is carried by photometry and integrated spectroscopy, together with stellar population synthesis models.

Some of them exploit limited parts of the optical spectrum, such as spectral indicators (e.g. the $4000-$\AA \ break, the \Hd$_A$ index, emission lines) combined with the photometric datum. \cite{kauffmann03} (hereafter K03) follow such kind of approach. They (but see also \citealt{tremonti04}) build a set of template spectra spanning various Star Formation Histories, extinction amounts and metallicities, for which the spectral indicators are computed and compared with the observed ones.

Other works, e.g. \cite{cidfernandes05} and more recently \cite{ocvirk06} and \cite{mathis06}, use the information carried by the entire optical spectrum. The model presented by \cite{cidfernandes05} searches the best combination of metallicity, overall extinction and Star Formation Rate values --at given epochs-- that minimizes the differences between observed and model spectrum. Besides stellar masses, star formation rates and extinction, they claim that in such a way it is possible to recover also an estimate of metallicity with a certain degree of accuracy.

The {\sc moped} algorithm, already used by \cite{heavens00} to analyse galactic spectra, has been used by \cite{mathis06} to reproduce the analysis of \sdss \ data performed by K03. \cite{ocvirk06} focus on the recovery of the star formation history and on the age--metallicity relation, by identifying the spectral features that carry the information one is looking for (i.e. age, metallicity, etc.).

In order to reduce the number of parameters and to simplify the analysis task, all the models reviewed thus far rely on the following hypothesis: 1) dust extinction is assumed to equally affect the stellar populations of all ages 2) the extinction law is modeled as a power-law or, alternatively, is taken to be the same as the one found for the solar system neighborhood (see e.g. \citealt{cardelli89}); 3) the chemical evolution issue is usually dealt with by using a given --small-- number of SSP set with different values of metallicity.

Furthermore, the present model reproduces simultaneously not only the stellar features, but also the nebular emission which comes out in form of emission lines in the youngest stellar populations, and is crucial to derive the very recent ($\leq 2\times 10^7$ yr) Star Formation Rate. This approach allows us to take into account the effects of both extinction and of absorption profiles on the lines intensity.

This work rests on the framework of the \wings \ project (see e.g. \citealt{fasano02} and \citealt{fasano06}), a survey which is providing multi-band photometry and spectroscopy of a sample of cluster galaxies in the nearby universe (redshift between $0.04$ and $0.07$), with the goal of setting a statistically robust local reference of the properties of cluster galaxies and their variances, prior to draw any conclusion about their evolution.

The aim of the spectroscopic campaign carried out on a \wings \ subsample is to shed new light on the link between the evolution of star formation and galaxy morphology, as well as the dependence on the characteristics of the cluster where they are found and the position within the cluster. Furthermore, the availability of data on low redshift clusters can be used as a reference for studies on distant clusters (see e.g. \citealt{poggianti06}). 

In order to analyse the spectral features and the history of star formation in cluster galaxies of the \wings \ sample, we developed and improved a pre-existent spectral fitting code that is capable of reproducing the galactic emission from the far UV to the Near--Infrared, including both the stellar and nebular emission. As for other similar models, it will provide values both for the overall stellar mass and for the mass of stars formed at given epochs of a galaxy's life, and hints on the metallicity for some cases. Furthermore, exploiting a new method of automatic measurement of equivalent width of spectral lines, we will also provide catalogs for the most important lines.

We first explain the main features of our model in Section {\bf 2}. In Section {\bf 3} we describe the automatic measurement of equivalent width of lines and its uncertainties. In Section {\bf 4} we give an example of application of our model on two observed spectra, so that the reader can easily become familiar with the method. In Section {\bf 5} we describe the test performed to verify the degree of accuracy of our method both on synthetic template spectra and on the observed ones. Finally, in Section {\bf 6} we discuss our method, its uncertainties and the results that one can reasonably obtain.

In this work we adopt a Friedmann--Robertson--Walker cosmology with $\Omega=0.3$, $\Lambda=0.7$ and H$_0=75$ km s$^{-1}$ Mpc$^{-1}$.

\section{THE MODEL}\label{sec:model}
The model described here is a significantly improved and extended version of the spectrophotometric code developed by \cite{poggianti01} to derive the star formation histories of a sample of Luminous Infrared galaxies from their average spectrum.  After minor improvements, the code was used by \cite{berta03} to perform a spatially resolved analysis of long-slit spectra of the Ultra Luminous Infra-Red Galaxy IRAS19254-7245, one of the best examples of a galaxy--galaxy ongoing collision in the nearby universe, also known as the {\it SuperAntennae}.

The goal of this model is to reconstruct the star formation history (SFH, hereafter) of galaxies, hence the amount of stars formed at each epoch throughout the galaxy evolutionary history. To this aim, the code reproduces the main features of an observed spectrum: the equivalent widths of several lines --both in absorption and in emission-- and the fluxes emitted in given bands of the continuum. The spectral region can extend from the UV to the near--infrared and the model can be considered reliable up to $\sim 5$ $\mu m$, while at longer wavelengths a treatment of the thermal dust emission component --either from the interstellar medium or from star forming regions-- would be required. A galaxy model spectrum is computed by adding the synthetic spectra of Single Stellar Populations (SSPs) of different ages built with a Salpeter initial mass function (IMF) with stellar masses in the range $0.15\le M \le 120$  M$_\odot$. Before being added, each spectrum is first of all multiplied by an appropriate mass value according to the SFH we aim to reproduce, then it is extinguished to simulate dust in a uniform screen geometry. The final synthetic spectrum will hence have the form:
\begin{equation}\label{eqn:spec}
L_{mod}(\lambda)=\sum_{i=1}^{N_{SSP}}M_i\cdot L_i(\lambda)\cdot 10^{-0.4\cdot A(\lambda)\cdot R_V \cdot E(B-V)_i}
\end{equation}
where $N_{SSP}$ is the number of SSP used, $M_i$ and $L_i(\lambda)$ are the value of stellar mass and luminosity of the {\it i-th} SSP respectively, $A(\lambda)$ represents the extinction law, $E(B-V)_i$ is the color excess assigned to the {\it i-th} SSP and finally $R_V$ is the ratio of total to selective absorption at V [$A_V/E(B-V)$]. Note that only positive values for extinction are allowed.

The most salient characteristics of this spectrophotometric model concerns the treatment of dust obscuration and the method for finding the best-fitting SFH and its related uncertainty. For the sake of generality we use the Galactic extinction curve ($R_V=3.1$, \citealt{cardelli89}), but the value of the color excess, $E(B-V)$, is allowed to vary from one population to another, according to the hypothesis of selective extinction: the youngest stellar populations are expected to be found at least partially nested in the dust of the molecular clouds where they were born, thus generally dust extinction will be higher for younger stellar populations \citep{calzetti94,powu,poggianti01}. Models that consider this factor include, for example, \cite{silva98} and \cite{charlot00}, which take into account the fact that birth clouds have finite lifetimes.

This treatment of dust obscuration is not only a more realistic description of the real situation --instead of assuming a common amount of extinction for stars of all ages--, but it is also desirable when, such as in our case, one wants to reproduce also the intensity of emission lines, which are the most prominent feature of the youngest stars, the more prone to dust extinction.

In fact, the typical difference in the obscuration amount as computed from emission line --hence affecting young stars-- and that computed from colours (see e.g. K03) --and hence affecting the continuum emission which is dominated by old and intermediate-age stars--, is about a factor of $2$. Assuming a common extinction value for all the stars in a galaxy may yield a good fit of its optical spectrum, but is a simplistic and a known-to-be-incorrect assumption.\\ 

\subsection{The set of SSP spectra}\label{sec:ssp}

The model we describe here can in principle work with any set of SSPs. The synthetic spectra used for the \wings \ project are taken from a combination of two different sets that both use Padova evolutionary tracks  \citep{bertelli94}. The first set uses the \cite{jacoby84} library of observed stellar spectra in the optical domain, and is composed of spectra of $108$ different ages, from $10^5$ to $20\times 10^9$ years that have been computed for five different values of the metallicity (namely $Z=0.05, 0.02, 0.008, 0.004$ and $0.0004$). The emission from ionized gas in the interstellar medium has been computed by means of the photoionisation code {\sc cloudy} \citep{ferland96}, so that the youngest SSPs display hydrogen and forbidden emission lines, mainly of oxygen, nitrogen and sulfur. The nebular emission has been computed assuming case B recombination \citep{osterbrock89}, an electron temperature of $10^4$ K and an electron density of $100$ cm$^{-3}$, the radius of the ionizing cluster is $R=15$ pc and its mass $10^4$ M$_\odot$, which were chosen as typical values in {\sc Hii} regions (A. Bressan, private communication). The spectral resolution in the optical --i.e. from $\sim 3500$ to $\sim 7500$ \AA-- is $FWHM=4.5$ \AA, while the spectra have been extended with Kurucz theoretical libraries to a wider wavelength range ($90-10^9$ \AA) at a much lower resolution. Furthermore, dust emission from circumstellar envelopes of AGB stars has been added as in \cite{bressan98}.

The second set of SSPs uses the MILES library, a more recent and larger library of observed stellar spectra \citep{Psanchez04,Psanchez06}. SSP spectra computed with this new library were kindly provided to us by Patricia S{\'a}nchez-Bl{\'a}zquez and collaborators. The main differences with respect to the previous set are:
\begin{itemize}
\item these SSPs are computed in the spectral range $3536\div 7412$ \AA
\item the spectral resolution is higher: $FWHM=2.3$ \AA \ instead of $4.5$\AA
\item the age range goes from $1$ to $17.78$ Gyr, so no population younger than 1 Gyr is included, for a total number of $26$ SSP
\item the values of metallicity for which they have been computed are (expressed in terms of $\left[Fe/H\right]$): $0.20, 0.00, -0.38, -0.68, -1.28$ and $-1.68$ 
\end{itemize}
We will, hence, use a combination of these two sets adequately adapted to our purposes. 

First, we reduced the spectral resolution of the two sets of SSPs in order to match the resolution of the observed spectra, which in the case of the \wings \ survey is $FWHM=6$ and $9$ \AA \ for observations in the northern and southern hemisphere, respectively.
To do this, the synthetic spectra are convolved with a gaussian whose $FWHM$ is calculated as:
\begin{equation}
FWHM_G=\sqrt{FWHM_{obs}^2-FWHM_{SSP}^2}
\end{equation}
In the range of wavelengths and ages where they are available we used the second set of SSPs, while the first set of SSPs of the same age was adopted at all other wavelengths, and for ages younger than $1$ Gyr. When the value of age did not have an exact correspondence between the two sets, we used the age-weighted average of the immediately younger and immediately older SSP. In all cases the final set of SSPs is normalized so that the bolometric luminosity is the same as for the first set. SSPs of similar ages, but belonging to the different sets, turned out to have negligible differences both in the colours and in the values of the equivalent widths of the main spectral lines. In this way we obtained a total of $108$ SSP spectra for three different values of metallicity, namely $Z=0.004$, $Z=0.020$ and $Z=0.050$.

One of the main hypotheses of our model is that all the stellar populations in a galaxy have a common metallicity value at any age. To avoid further complications that could arise when trying to account for the chemical evolution of galaxies, we decided to follow \cite{tremonti04} (but see also \citealt{mathis06}) and obtain --for each galaxy-- $3$ models using SSPs of different metallicities: sub-solar ($Z=0.004$), solar ($Z=0.02$) and super-solar ($Z=0.05$). We will choose hence, as the best fit model, the one among the different metallicity models, with the smaller $\chi^2$ (see also \S \ \ref{sec:met} for a more detailed discussion). We are in this way interpreting the value of metallicity we found as belonging to the stars dominating the light in a spectrum.

\subsubsection{Reducing the number of SSPs}\label{sec:sspeqw}

A fine age grid of SSPs such as the one described above is much beyond the needs of our work and would only increase the number of parameters of the spectral fitting. In fact, a very fine grid in ages is redundant and does not increase the power of the model given the intrinsically very small differences in the spectra of stars of only slightly different ages. In previous works (i.e. \citealt{poggianti01} and \citealt{berta03}) we used just $10$ SSPs: $4$ young ones (i.e. with emission lines), $5$ of ``intermediate-age'' (i.e. with prominent Balmer lines in absorption and calcium lines) and $1$ ``old'' population composed of SSPs older than $10^9$ years, which were summed together in order to mimic a constant SFR over $\sim 10^{10}$ years. Each SSP was considered as representative of the emission of stars born within suitable age ranges centered on the SSP's age.

\begin{figure*}
\begin{center}
\rotatebox{-90}{
\includegraphics[height=1.0\textwidth]{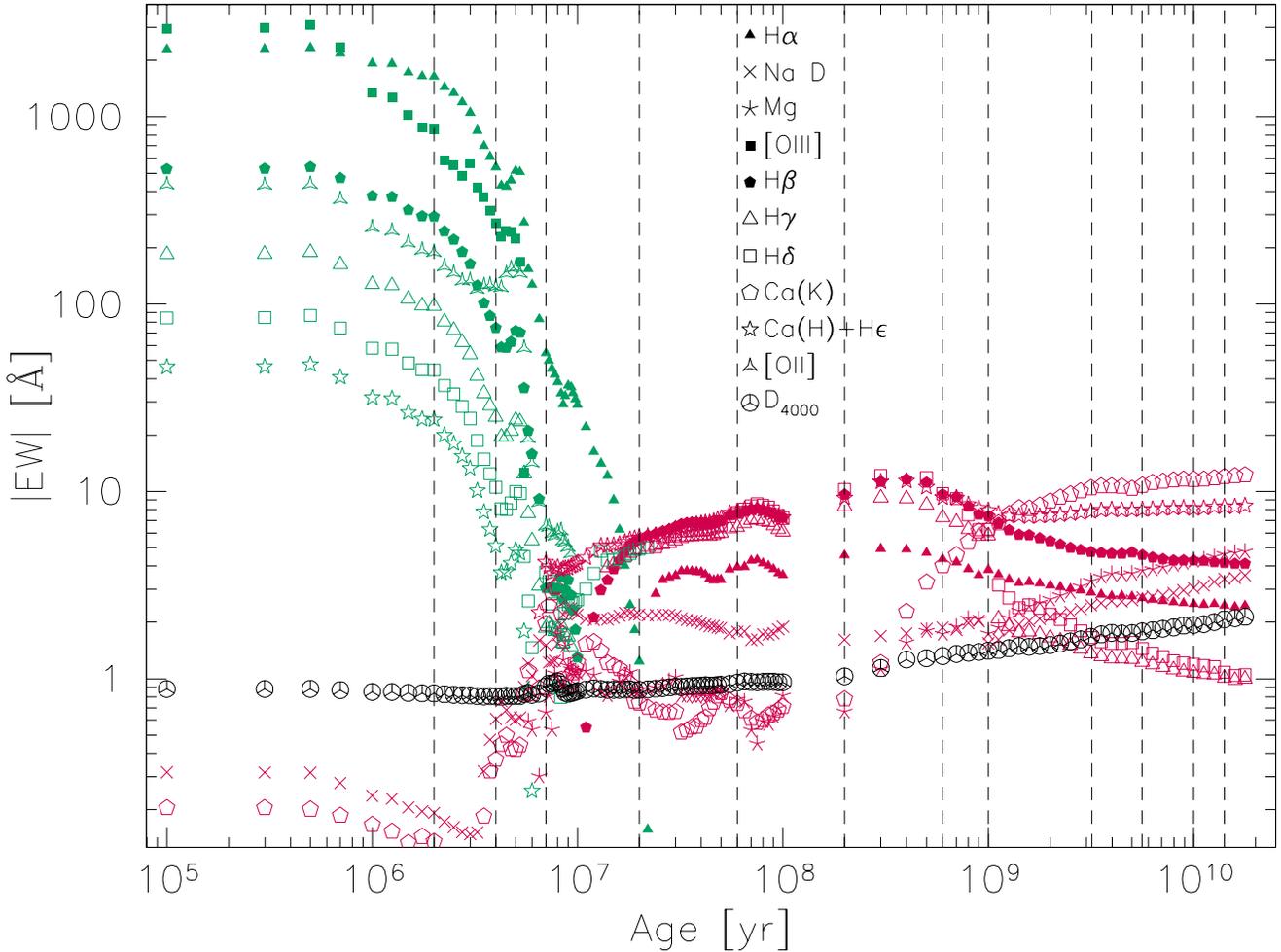}}
\caption{Values of equivalent width of lines and of the D$_{4000}$ index in the complete set of SSP spectra adopted and described in \S \ref{sec:ssp}. All are measured with the method described in section \ref{sec:eqw} in spectra that have been degraded to a resolution $FWHM=6$\AA. Note that equivalent widths are expressed in \AA, and for convenience, their absolute values are plotted. Green points refer to lines measured in emission (negative value of EW), while red ones are measured in absorption. Vertical dashed lines identify the time intervals of the 13 averaged SSPs as in Table 1.}
\label{fig:ssp}
\end{center}
\end{figure*}

Here we decided to follow a slightly different approach. We built a new set of stellar population spectra starting from the $108$ SSPs, averaging all the SSP spectra with ages within given time intervals. This effectively simulates a constant star formation rate during each interval\footnote{It is worth stressing here that the spectra built in this way cannot be considered as belonging to Simple Stellar Populations anymore. For simplicity we will continue to use this acronym even when speaking of the new set of ``averaged'' SSPs.}. 
The time intervals were chosen on a physical basis, considering the stellar ages at which the main spectral features change significantly. In particular, we relied on the following features:
\begin{itemize}
\item the value of the equivalent widths of the following lines: Balmer lines, calcium lines ({\sc k} and {\sc h}), \Oii \ and  \Oiii
\item the value of the D$_{4000}$ index
\end{itemize}
Figure \ref{fig:ssp} illustrates the trend of the features mentioned above and how the choice for the ages of the bins was made. Furthermore we tried to take into account the characteristic evolutionary time of stars, so that stars in a given phase of their evolution are not split in different bins but rather contained within a single one. For instance, the third bin in Fig. \ref{fig:ssp} includes the stars in the Wolf-Rayet phase, which have the hardest emission (e.g. UV continuum) and dominate the SSP light between $4$ and $6$ Myr, while the fourth bin includes stars in the Super-Red Giant phase.
Spectral line characteristics as a function of age will be exploited as the most reliable way to assign an age value to the dominant stellar populations, while the characteristics of the emission in the continuum will constrain both the total stellar mass and the dust attenuation.

Note that, due to the logarithmic scale of Fig. \ref{fig:ssp}, the values of equivalent width of the older SSPs seem to early reach an almost constant trend for ages older than $\sim 2\times 10^8$ years. However, when measured on the $13$ age--binned spectra, the spectral lines show remarkably differences from one age to another.

Table \ref{tab:ssp} spells out the ages and the time interval $\Delta T$ over which we consider a given population to have a constant star formation rate. As arbitrary naming convention of these ``average SSPs'', in the following we refer to the age of the oldest stellar population in that time interval.
\begin{table}
\centering
\begin{tabular}{r l | c}
Age interval     &  [yr]             & $\Delta T$ [yr]  \\
\hline
$(0\div 2)      $&$\times 10^6	 $   & $ 2.0 \times 10^6$ \\
$(2\div 4)      $&$\times 10^6	 $   & $ 2.0 \times 10^6$ \\
$(4\div 7)      $&$\times 10^6	 $   & $ 3.0 \times 10^6$ \\ 
$(7\div 20)     $&$\times 10^6	 $   & $ 1.3 \times 10^7$ \\
$(2\div 6)      $&$\times 10^7	 $   & $ 4.0 \times 10^7$ \\ 
$(6\div 20)     $&$\times 10^7	 $   & $ 1.4 \times 10^8$ \\ 
$(2\div 6)      $&$\times 10^8	 $   & $ 4.0 \times 10^8$ \\	    
$(6\div 10)     $&$\times 10^8	 $   & $ 4.0 \times 10^8$ \\ 
$(1\div 3.2)    $&$\times 10^9	 $   & $ 2.2 \times 10^9$ \\
$(3.2\div 5.6)  $&$\times 10^9	 $   & $ 2.4 \times 10^9$ \\
$(5.6\div 10)   $&$\times 10^9	 $   & $ 4.4 \times 10^9$ \\
$(1\div 1.41)   $&$\times 10^{10}$   & $ 4.1 \times 10^9$ \\
$(1.41\div 1.78)$&$\times 10^{10}$   & $ 4.1 \times 10^9$ \\
\hline
\end{tabular}
\caption{The ages and durations ($\Delta T$) of the set of averaged SSP spectra used in this work, as it was built according to the criteria explained in the text and in figure \ref{fig:ssp}. Here $\Delta T$ is the time interval over which the SFR is assumed to be constant.}
\label{tab:ssp}
\end{table}
In total, we use $13$ averaged SSP spectra of fixed metallicity. 

\subsection{The Best-fitting and Optimization Algorithm}\label{sec:simann}

Given the SSP spectra and the extinction law, the space parameter is defined by the $13$  values of mass and $13$ of extinction, one for each stellar population. The code compares each observed spectrum, previously corrected in order to account for our Galaxy extinction using \cite{schlegel98} extinction maps, with synthetic spectra automatically performing a random exploration of the parameter space to find the best combination of mass and extinction values that reproduces the observed spectrum. The comparison between model and observed spectrum is done on the flux measured on a certain number of bands in the continuum and on the values of equivalent width (hereafter EW) of spectral lines both in emission and in absorption (see table \ref{tab:bands} for the definition of continuum bands and for the list of EWs). The best fit to the observed spectrum is then searched by means of the minimization of a $\chi^2$ function, which is given by:
\begin{equation}\label{eqn:mf}
\chi^2=\sum_{i=1}^N \left(\frac{M_i-O_i}{E_i}\times W_i \right)^2 \slash \sum_{i=1}^N
\end{equation}
in which $N$ is the total number of spectral features used to constrain the model, $M_i$ and $O_i$ are respectively the model and the observed values of continuum and line emission, $E_i$ the observational error.

Due to the presence of extinction, finding the set of values that better reproduces an observed spectrum turns out to be a non-linear problem defined in a $S=2\times N_{SSP}$--parameter space, where $N_{SSP}$ is the number of synthetic SSP spectra adopted. Moreover, issues like the one we are dealing with here are characterized by the presence of lots of local, non--absolute, minima. To achieve the best result we used the Adaptive Simulated Annealing algorithm (ASA, \citealt{Ing1} and \citealt{ASA}), which is particularly well suited to find solutions for such problems, being able to converge to an absolute minimum and to avoid resting in points of local minimum. We defer to Appendix A for a more detailed description of the ASA algorithm.

\begin{table}
\centering
\begin{tabular}{c c|c c c c l l}
&$\lambda_i$ & $\lambda_f$ &  &  &  & {\sc line } & $\lambda_C$ \\
\cline{2-3}\cline{7-8}
&2370 & 2580 &  &   &  &H$\alpha$		  & $6563^\star$ \\
\cline{2-3}\cline{7-8}	  				
&2940 & 3200 &  &   &  & Na ({\sc d})	          & $5890+5895$ \\
\cline{2-3}\cline{7-8} 
&3500 & 3640 &  &  &   & Mg			  & $5177$ \\
\cline{2-3}\cline{7-8}	      
&3750 & 3950 &  &  &   & [O{\sc iii}]		  & $5007$ \\
\cline{2-3}\cline{7-8} 
&3850 & 3950 &  &  &   &H$\beta$		  & $4861^\star$ \\
\cline{2-3}\cline{7-8} 
&4000 & 4100 &  &  &   &H$\gamma$		  & $4341$ \\
\cline{2-3}\cline{7-8} 
&4050 & 4250 &  &  &   & CO {\sc g}-band	  & $4305$ \\
\cline{2-3}\cline{7-8} 
&4410 & 4510 &  &  &   &H$\delta$		  & $4101^\star$ \\
\cline{2-3}\cline{7-8} 
&4670 & 4820 &  &  &   &H$\epsilon$+Ca{\sc ii (h)}& $3969^\star$ \\
\cline{2-3}\cline{7-8}
&5050 & 5150 &  &  &   &Ca{\sc ii (k)}  	  & $3934^\star$ \\
\cline{2-3}\cline{7-8} 
&5250 & 5460 &  &  &   &H$\zeta$		  & $3889$ \\
\cline{2-3}\cline{7-8} 
&5770 & 5860 &  &  &   &H$\eta$ 		  & $3835^\star$ \\
\cline{2-3}\cline{7-8} 
&5950 & 6140 &  &  &   &H$\theta$		  & $3798$ \\
\cline{2-3}
&6380 & 6480 &  &  &   &[O{\sc ii}]		  & $3727^\star$ \\
\cline{2-3} 
&6960 & 7120 &  &  &   &			  &  \\
\cline{2-3} 
&7410 & 7570 &  &  &   &			  &  \\
\end{tabular}
\caption{Here we report, on the left, the wavelengths expressed in \AA, of the boundaries for the bands that are used to constrain continuum emission. On the right, the list of spectral lines that are measured. Only lines marked with a $^\star$ are effectively used as constraints in the fit.}
\label{tab:bands}
\end{table}

The CPU time required to fit a spectrum depends on parameters such as the number of SSP adopted, the number of constraints used and the accuracy taken to explore the whole parameter space. From the tests that we performed we found that a good result, in terms of values of the merit function, was achieved in about $20$ seconds for each spectrum on an Intel Xeon --3.06 GHz-- CPU.

The observational errors that are used in the function defined by Eq. \ref{eqn:mf} are computed as follows: for the equivalent widths of the spectral lines a ``poissonian'' error is used, so that:
\begin{equation}\label{eqn:lnserr}
E_{lns}=0.5 \cdot \sqrt{|EW|}+E_M
\end{equation}
where $E_M$ represents an estimate of the uncertainty of the method used to measure the line (see \S \ref{sec:eqw}). A minimum error is also set ($\sim 1$ \AA \ for the \wings \ dataset) in case the one given by equation \ref{eqn:lnserr} is too low. 

Computing the error on the continuum flux is a non-straightforward task, mainly because the signal-to-noise ratio (S/N) is not constant over the spectrum but can vary significantly depending on wavelength. The error on the continuum flux is computed, for each continuum band, as follows: an average S/N value is defined within a given wavelength interval, and the error is obtained as the mean square root with respect to the average value. 
Since the continuum slope can change within a given band, and since we want to avoid that this fact is interpreted as a source of noise, this is done over sufficiently small wavelengths intervals, dividing each continuum band in a given number of parts which can be adequately set according to the S/N ratio. Given the characteristics of the bands and of our observed spectra, we fixed its value to $4$. The observed error for any given continuum band is then computed as the arithmetic average of the $\sigma$ computed on each of these parts. 

The position and broadness of the continuum bands to consider can be chosen in two different ways:
\begin{enumerate}
\item one by one for each observed spectrum
\item fixed bands for all the spectra
\end{enumerate}
While the first option would be the most accurate one, it is unrealistic to use when the number of spectra is high ($>100$). With the second option a certain number of bands is suitably defined in a $z=0$ template spectrum. These bands are red--shifted to the rest-frame of the objects, and only those that lie within the observed spectra wavelength range are used. 

When using the second option, as we do for the \wings \ dataset, the choice of the continuum bands was made using as reference three synthetic spectra explicitly build so to represent typical galactic spectra: a passively-evolving galaxy (to mimic galaxies dominated by old stars), a post-starburst galaxy dominated by deep Balmer absorption lines and finally an actively star-forming spectrum, with strong emission lines and blue continuum. The bands, for each spectrum, were chosen so that none would overlap spectral lines and in such a way to sample the slope of the continuum in the best possible way. Finally a compromise between the three sets was used to build the final list which we report in table \ref{tab:bands}. In order to adequately constrain the continuum emission of galaxies in the whole redshift range of the \wings \ spectra, some ultraviolet bands were added so that spectra with redshift up to $z\sim 0.6$ are suitably covered.

Due to relative-flux calibration issues (see Cava et al., in preparation) and the sky contamination in the red, the ``outer'' spectral regions of \wings \ spectra are often not reliable. For this reason, they were systematically cut by fixed amounts --$250$ and $500$ \AA \ respectively on the ultraviolet and on the red side-- to avoid a misleading interpretation of the spectral features.

In table \ref{tab:bands} we also summarize the spectral lines that are measured in the observed spectra as explained in \S\ref{sec:eqw}. Not all the lines shown are used to constrain the model fit. For example, the intensity of the {\sc [Oiii]} ($5007$ \AA) line is hard to model because it is very sensitive to the physical condition of the gas where it is emitted, while the Mg line at $5177$ \AA \ is sensitive to the enhancement of $\alpha-$elements which is not included in our SSPs, hence they are not considered.

\subsection{Other improvements in the model}

Given the set of SSP spectra used, the version of the code described here deals mainly with spectral features and continuum in the range $3500-7500$ \AA \ rest frame, but the spectral analysis can be in principle extended also to other wavelengths although with less detail, i.e. at lower spectral resolution.
To exploit this feature, we introduced the possibility of using also photometric data (e.g. observed magnitudes or broad--band fluxes). The spectral range of photometric data that can be used is only limited by the reliability of the synthetic spectra that is, with this set of SSP, from the far-UV to $\sim 5$ $\mu$ (see also \S \ref{sec:model}). With this addition it is possible to combine spectral and photometric information, and it is particularly useful to constrain near-infrared (e.g. J, H, K bands) or ultraviolet emission. To use such information, the observed magnitude is rescaled with the same normalization as for the observed spectrum (see Eq. \ref{eqn:resc}). The flux value of the model spectrum in a given band is obtained as the integral convolution of the spectrum itself with the response curve of the filter.

Also note that, when comparing an observed spectrum with a synthetic one, the model is redshifted to the galaxy rest-frame. Such approach is useful when also broad-band data are used to constrain the model, avoiding in this way the introduction of a model dependent K-correction on the observed data.

When dealing with spectra with an absolute flux calibration, the analysis performed by means of this model will yield the total mass of stars and the Star Formation Rate --along the evolutionary phases of the galaxy-- computed as the ratio between the mass of stars in a given time interval and the time interval itself.  

The fit is performed over the observed spectrum whose flux is normalized to $1$ at a given $\lambda_{ref}$ (usually at $5550$ \AA) and so is the model. The SFR values, referring to such a normalization, have then an arbitrary scale. Since the SSP spectra are all given in unit of $10^{30} erg/s/$\AA$/M_\odot$, to obtain the ``physical'' values for SFR and stellar masses, they must be re-scaled by a normalization factor, $C_N$:
\begin{equation}\label{eqn:resc}
C_N=4 \pi d_L^2 \cdot \frac{F_{obs}(\lambda_{ref})}{10^{30}\cdot L_{MOD}(\lambda_{ref})}
\end{equation}
where $F_{obs}(\lambda_{ref})$ is the observed flux, or magnitude, at the reference wavelength $\lambda_{ref}$, $L_{MOD}(\lambda_{ref})$ is the luminosity of the model at the same wavelength and $d_L^2$ is the luminosity distance. This yields physical values of masses in solar masses and SFR in solar masses per year. Since the luminosity of each SSP spectrum refers to $1 \; M_\odot$ of initial mass, a correction for the actual mass value of the SSP must be done in order to correctly take into account the mass losses due to the stellar evolution. This correction is performed by multiplying the stellar mass by the mass fraction of live stars, which is a function of age and, for the oldest stars, can be as high as $\sim 65$ per cent.

If the spectra do not carry an absolute flux calibration, the V-band magnitude --measured in the same aperture as spectroscopy-- is used to re-scale them to an observed flux. In this way, the values of stellar masses and star formation rates obtained, refer to that given aperture. Furthermore, an estimate for the stellar mass of the galaxy as a whole is obtained using the total V-band magnitude, assuming the stellar populations sampled by the spectroscopy are representative of the whole galaxy.

The model yields also a value for extinction in the $V$ band, which is computed as follows 
\begin{equation}\label{Av}
A_{5550}=-2.5 \times \log_{10}\left[\frac{L_{tot}^M(5550)}{L_{unext}^M(5550)}\right] \times A(5550)
\end{equation}
where $L_{tot}^M$ and $L_{unext}^M$ are the best--fit model spectrum and the model spectrum obtained setting $A_{Vi}=0$ for each SSP respectively, and $A_V(5550)$ is the value of extinction curve calculated at $\lambda=5550$ \AA. This value in particular is the overall extinction, i.e. a weighted average of every population's extinction. To compare the model value with the one obtained with standard techniques, i.e. Balmer emission lines ratio, one should take into account only SSPs that display these lines --typically \Ha \ and \Hb--, that are those with age less than $2\cdot 10^7$ years.

If the observed far infrared (hereafter FIR) luminosity is also known, this can be used as a further constraint. In the hypothesis that all the radiation that is attenuated in the model is thermally re-emitted by dust in the spectral range $\sim 10 \div 1000$ $\mu m$, the bolometric FIR luminosity of the model will be computed as the difference between UN-absorbed and absorbed spectrum:
\begin{equation}\label{fir}
L_{FIR}=4 \pi d_L^2 \times \int_0^{\infty} \left[F_{unext}^M(\lambda)-F_{tot}^M(\lambda)d\lambda \right]
\end{equation}
where $d_L$ is the luminosity distance of the object. 

This is an information of fundamental importance when dealing with galaxies with a strong starburst activity, since they can host a substantial quantity of young stars deeply buried inside the dense molecular clouds where they were just born. These stars can be almost totally invisible at optical wavelengths, their presence being revealed only in the infrared domain, where dust thermally re-emits their energy. The only way of correctly taking into account for this fact is the use of the observed infrared luminosity as further observational constraint. In such cases both extinction and SFR can be strongly underestimated if only the optical spectrum is taken into account. 

Since we expect the \wings \ sample to have only a few of such powerful infrared emitters and of galaxies with very high SFR and strong extinction, the maximum values allowed for extinction of the youngest SSP are kept within ``normal'' values i.e. those of quietly star-forming galaxies ($A_V < 1$).

The value of current SFR obtained with this spectral fitting is in principle very reliable: stars younger than $10^7$ years are the ones responsible for emission lines which are usually used as an estimator of current star formation activity. In order for such diagnostic to yield reliable values, line fluxes of Balmer lines, for example \Ha, besides being corrected for dust extinction, need to be corrected also for the presence of an absorption component originating in the stellar photospheres. The magnitude of this absorption component is not constant in general, but can vary according to the SFH of the galaxy. In any case this can affect the value of extinction when it is computed from the observed Balmer lines ratio. When this happens, the above ratio is usually over--estimated and so is the extinction, leading in turn to an excessively high value of SFR. Since in our model we measure the total effect of emission$+$absorption in the EW value, this bias should be naturally avoided.

Besides the emission coming from the stars, the luminosity of Seyfert galaxies can have a substantial contribution from a non-thermal source. In the optical domain such emission is detected mostly via emission lines, in the case of Seyfert-2, but also in the continuum emission in type-1 Seyfert. In fact, the optical spectrum of Seyfert-1 galaxies can be in general represented as a power-law emission dominated by iron lines. In the case of Seyferts 2 the optical spectrum is dominated by starlight and can be well reproduced by means of SSP spectra, so the presence of a non-thermal source can be determined only by means of standard diagnostic diagrams of high ionization line ratios. In the case of Seyfert-1 the spectrum cannot be adequately reproduced, simply because it is dominated by other mechanism of emission rather than thermal. Within our model, a Seyfert-1 galaxy is then recognizable because the $\chi^2$ of the model fits is unacceptable.

\section{EQUIVALENT WIDTH COMPUTATION}\label{sec:eqw}

\begin{figure*}[ht]
\centering
\begin{tabular}{l l}
\includegraphics[height=0.495\textwidth]{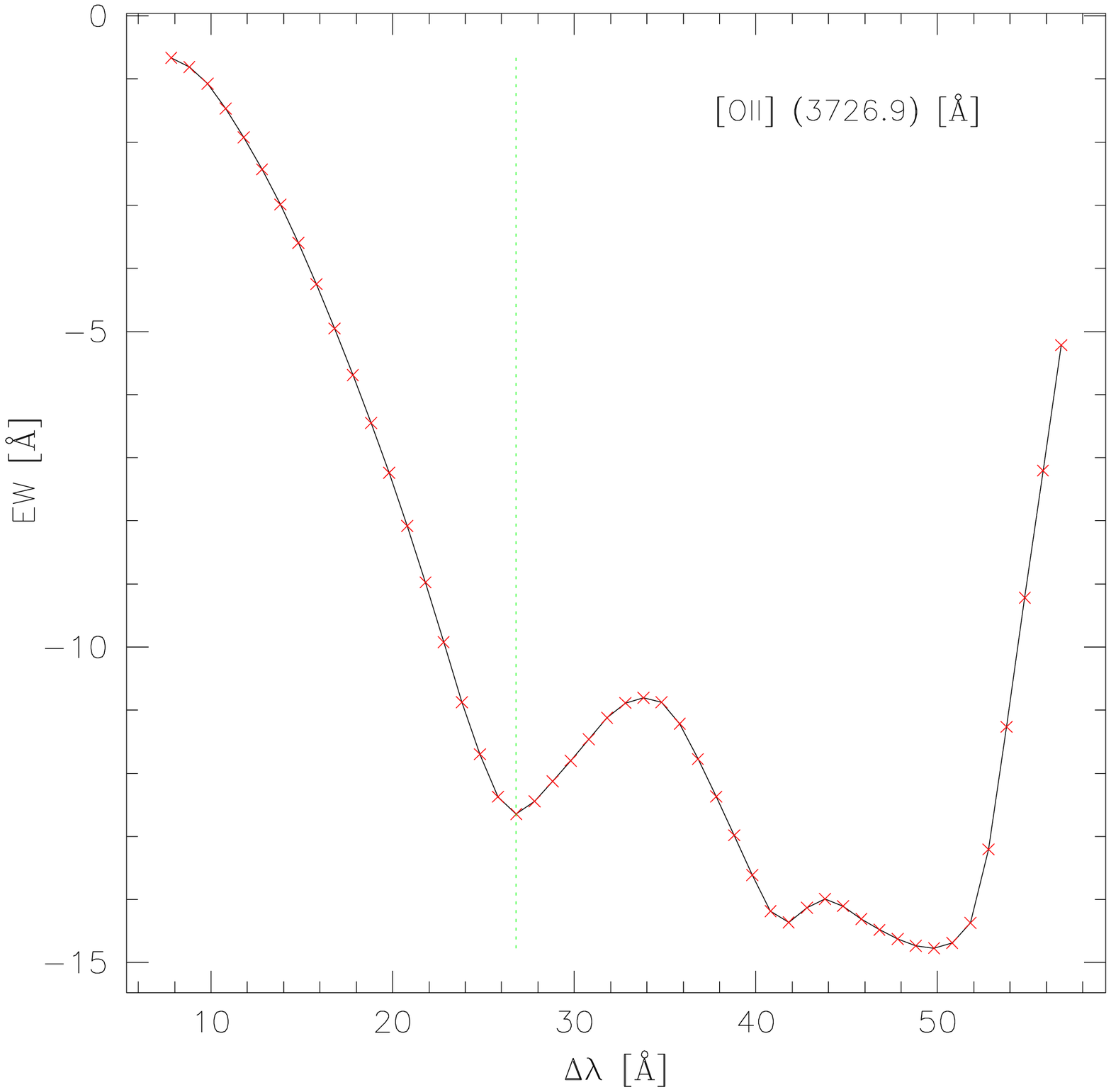} &
\includegraphics[height=0.495\textwidth]{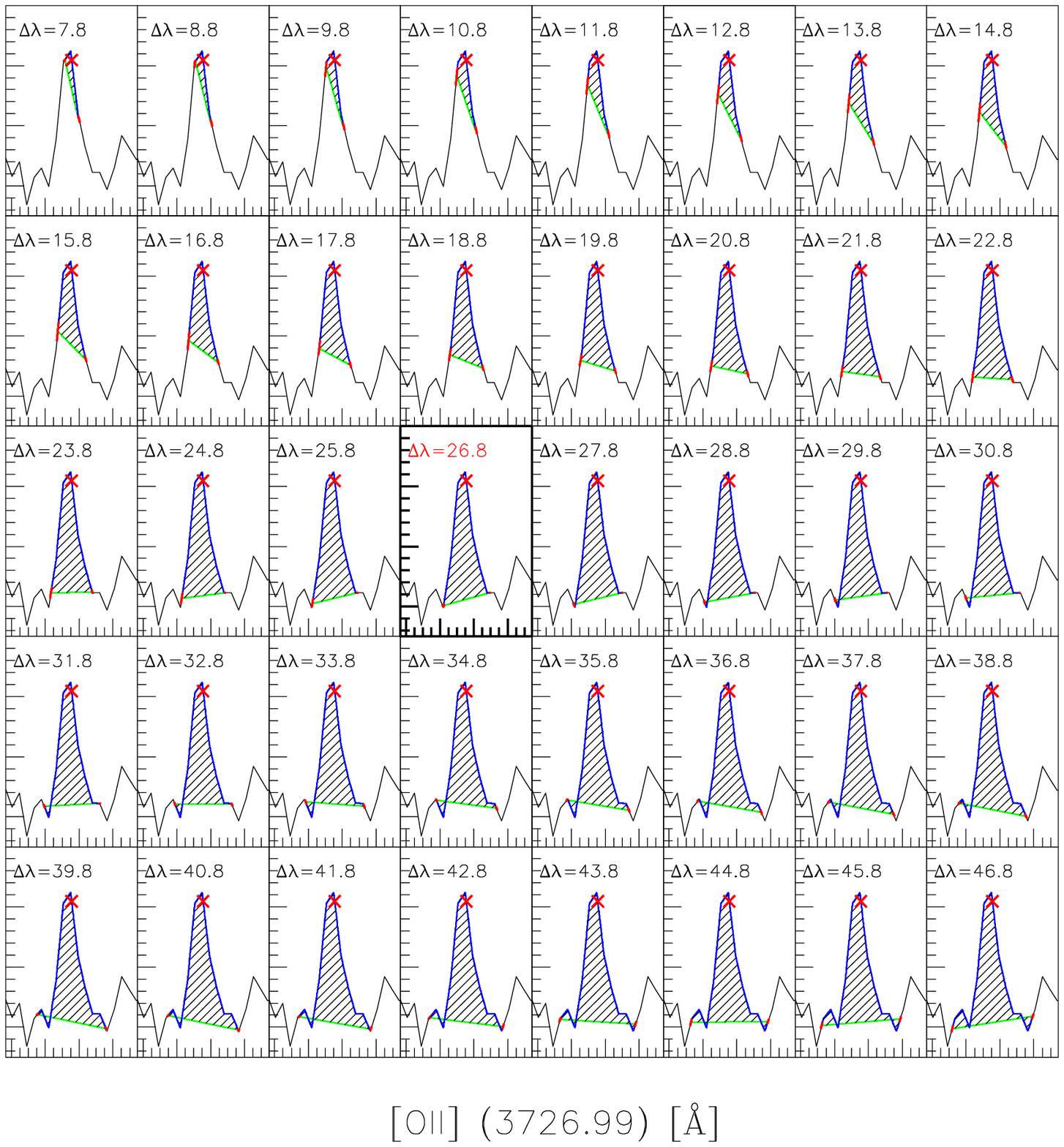} \\
\end{tabular}
\caption{Here we show an example of the procedure that we follow to compute line equivalent widths. This particular case shows the \Oii \  line at $3727$ \AA \ in an observed spectrum. In the left panel we show the value of EW as a function of $\Delta\lambda$: the red crosses represent the values for which the EW is computed (i.e. once every angstr\"om), while the vertical line shows the point of ``right $\Delta\lambda$ value''. The right panel displays the measurement of the line for the corresponding points of left panel (we show only the first $40$ measurements). In this case the line measurement is made at $\Delta\lambda\simeq 27$ \AA, where the first minimum of the trend function is found and, on the right panel, it is shown with a thicker line.}
\label{fig:eqwdet}
\end{figure*}

When trying to reconstruct the characteristics of the stellar populations in a galaxy, a very important information is carried by the spectral lines, and it is essential to measure their intensity with a good accuracy. Given the great number of observed spectra of the current spectroscopic surveys, including \wings, it is unfeasible to manually measure EW values of several lines in each spectrum and automatic measurement procedures are required.

We hence developed a technique which is capable of automatically measure the EW of lines both in emission and in absorption being also able to distinguish if the line is absent or if it is dominated by features such as bad sky subtraction or spectral noise.

The main issue when computing the EW of a line, both in emission or in absorption, is an appropriate choice of the continuum. We define $\Delta\lambda_l$, an interval centered on the wavelength of the line $\lambda_c$, that contains the whole line profile and whose extremes, $\lambda_c-\Delta\lambda_l$ and $\lambda_c+\Delta\lambda_l$ respectively, identify the extremes of the straight line that at the central wavelength of the line defines the continuum. Applying the definition of EW, the flux (emitted or absorbed) within the line is divided by the continuum value.

Following this approach, the problem of defining the spectral continuum translates into the correct choice of the quantity $\Delta\lambda$, which defines both the continuum and the line width. This in turn depends on various factors such as the spectral resolution and the velocity dispersion of stars or gas motions. 
To find the most appropriate value of $\Delta\lambda$, the EW is measured at increasing values of $\Delta\lambda$ and subsequently the EW trend curve is analyzed. To study the behavior of this trend curve we applied this method to a sample of low and mean signal-to-noise \wings \ spectra and to the set of $13$ SSP spectra described above, as representative of high signal-to-noise data. The main source of uncertainty for such a method is in general the presence of spikes typical of low S/N spectra, that may lead to a wrong choice of the continuum in the line's proximity, or may cause a line to be measured even when it is absent or dominated by noisy features. 

\begin{figure*}
\centering
\rotatebox{-90}{
\includegraphics[height=.95\textwidth]{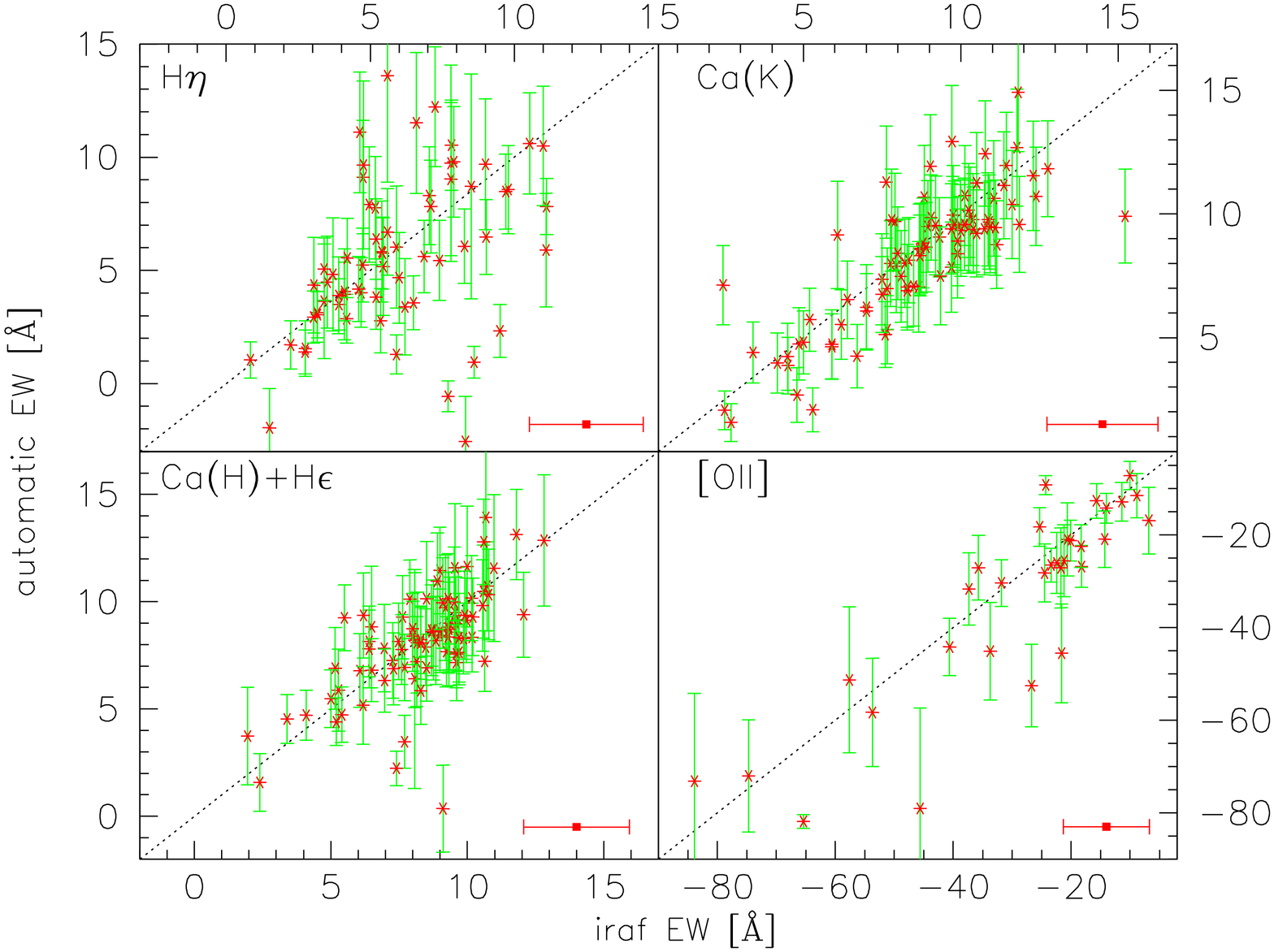}}
\caption{In this plot we show a comparison between the measurements of $4$ spectral lines --chosen among those listed in table \ref{tab:bands}-- made by using standard technique in {\sc iraf} (on the x axis) and by means of our automatic method. Note that to obtain the values of equivalent width no Gaussian fitting to the line profile was applied. The comparisons plotted here were chosen in order to give an idea of the reliability of our method, so here we present the case of the worst agreement --H$\eta$--, the two calcium lines which are of fundamental importance in the set of constraints, and the oxygen line as an example of an emission line. The comparison is made on $\sim 130$ spectra taken from the Abell clusters A119 and A3376, of the \wings \ sample. The EW values for these lines in the case that they were absent from a spectrum or not detected because below the noise level are not shown, but both automatic and {\sc iraf} measurements give similar detections. The horizontal, red error-bar on the lower-right of each panel, represents the typical uncertainty on the {\sc iraf} measurement.}
\label{fig:eqwcomp}
\end{figure*}
Adopting a fixed value, suitably chosen for each spectral line, of $\Delta\lambda$ would be --on one side-- a much more simple and straightforward approach. On the other hand it could lead to both a non-accurate measurement and to a misinterpretation of noisy features due, e.g., to bad sky subtraction or to the poor quality of a spectrum. Furthermore, if for a given spectral line a fixed value of $\Delta\lambda$ could be adequately used for the line measurement in a stellar population of a certain age, it will be not good for stellar populations of different ages as the strength of the line changes (see the discussion of this aspect in \citealt{dressler04}). This is particularly critical for the high order Balmer lines, such H$\epsilon$, which are the weakest, less defined and more crowded.

As it is expected, in general and for a good S/N spectrum, the EW tends to reach an asymptotic values as $\Delta\lambda$ increases, so for the great majority of cases, the first change of sign in the derivative of the trend curve is taken as the main measurement criterion for both absorption and emission lines (see Fig. \ref{fig:eqwdet}). Since in this way, the pattern ``emission+absorption'' --which is quite typical for \Hb \ e.g.-- would not be correctly detected, a check is also performed to highlight the second zero of the derivative function. This is only allowed for hydrogen lines from \Hb \ to H$\epsilon$ and only if the \Oii \ is detected in emission, i.e. when there are enough UV photons to ionise the interstellar gas.

A different treatment is reserved to the \Ha \ line. When this line is measured in emission, it is always accompanied by the two nitrogen lines at $6548$ and $6584$ \AA \ respectively. In medium resolution spectra, like those of the \wings \ sample, they often happen to be at least partially blended. Dealing with such an emission pattern by means of the method described above would imply the study of a very complicated trend curve. In this case we measure a first estimate of the EW, by adopting a fixed $\Delta\lambda=15$ \AA. This will allow to determine whether the line is detected in emission or in absorption with a good accuracy. In the first case the measurement is made by adopting a value of $\Delta\lambda$ of $70$\AA. In this way we measure also the emission coming from nitrogen, so that the value will include the three lines taken altogether. In case the line is detected in absorption we measure the line as done for the others.

The characteristics of the EW trend curve are also used to estimate the error that this method can, at least in principle, introduce to the EW estimate. When a line is measured in a high S/N spectrum, the value of the EW that we obtain by means of this method, reaches an asymptote since in such a spectrum the continuum level is well defined. If --on the contrary-- the S/N is low, this will introduce fluctuations in the EW measurements. We found that a good estimate is obtained by taking the semi-difference of the EW values measured with a $\Delta\lambda$ of $5$ \AA \ smaller and larger, respectively, than the value used for the line measurement. This, together with a factor that depends on the line intensity, is used to compute the error on EW.

In Fig. \ref{fig:eqwcomp} we show the comparison between the values of EW of four lines as measured one by one using the {\sc iraf} task {\tt splot} (x axis) and by means of our method. The case of H$\eta$ is shown because it has been found to yield the worst agreement between the two methods. The \Oii \ lines are shown as an example of emission line measurements (although there are few spectra displaying this line), while the two calcium lines are shown because they are particularly critical and important for the SFH reconstruction. These two lines in particular show a very good agreement with the ``manual'' measurement which is, anyway, done in a somewhat arbitrary manner.

\section{THE MODEL AT WORK}

\begin{figure*}
\centering
\begin{tabular}{l l}
\rotatebox{270}{\includegraphics[height=0.51\textwidth]{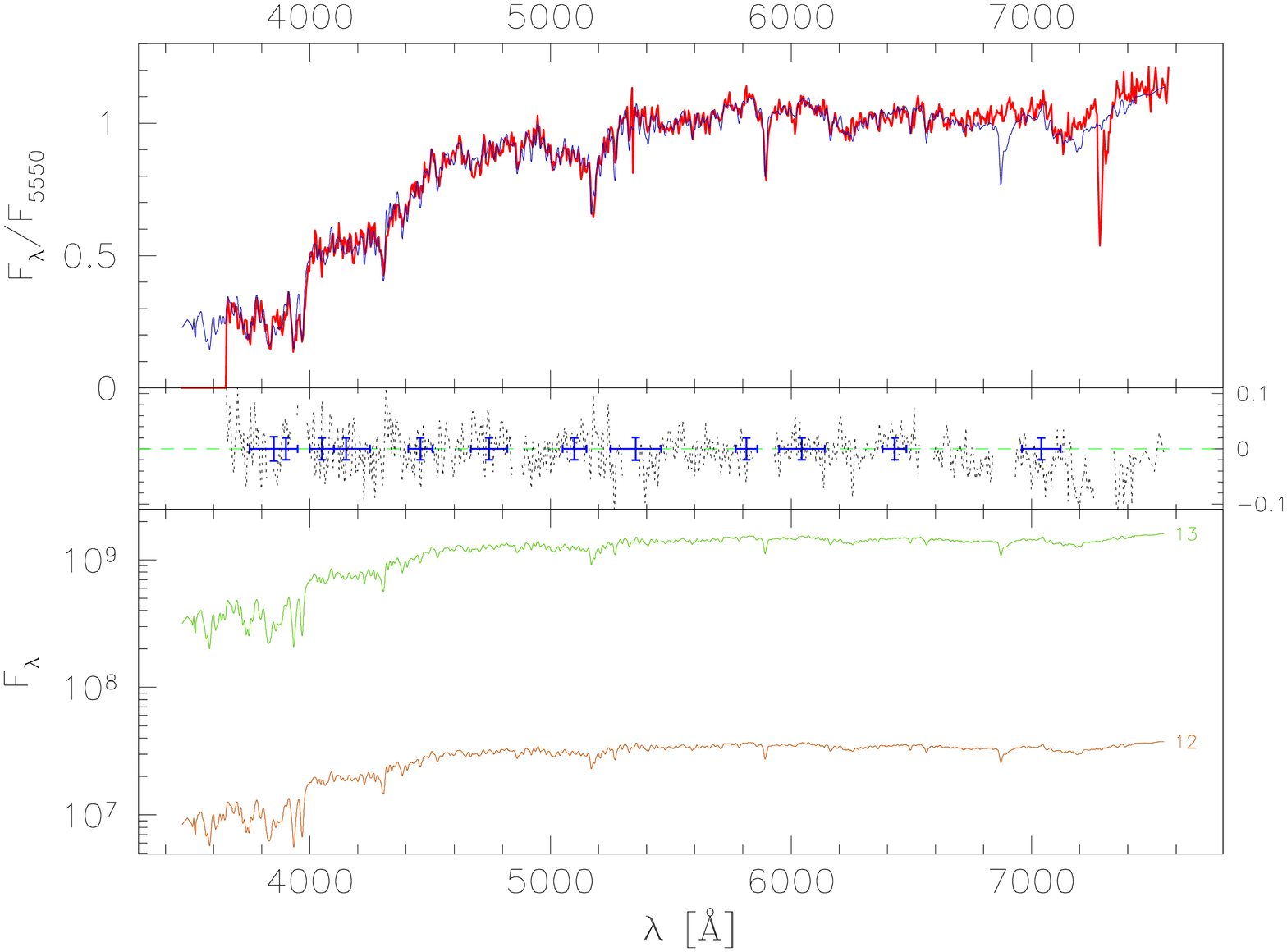}} &
\rotatebox{270}{\includegraphics[height=0.5\textwidth]{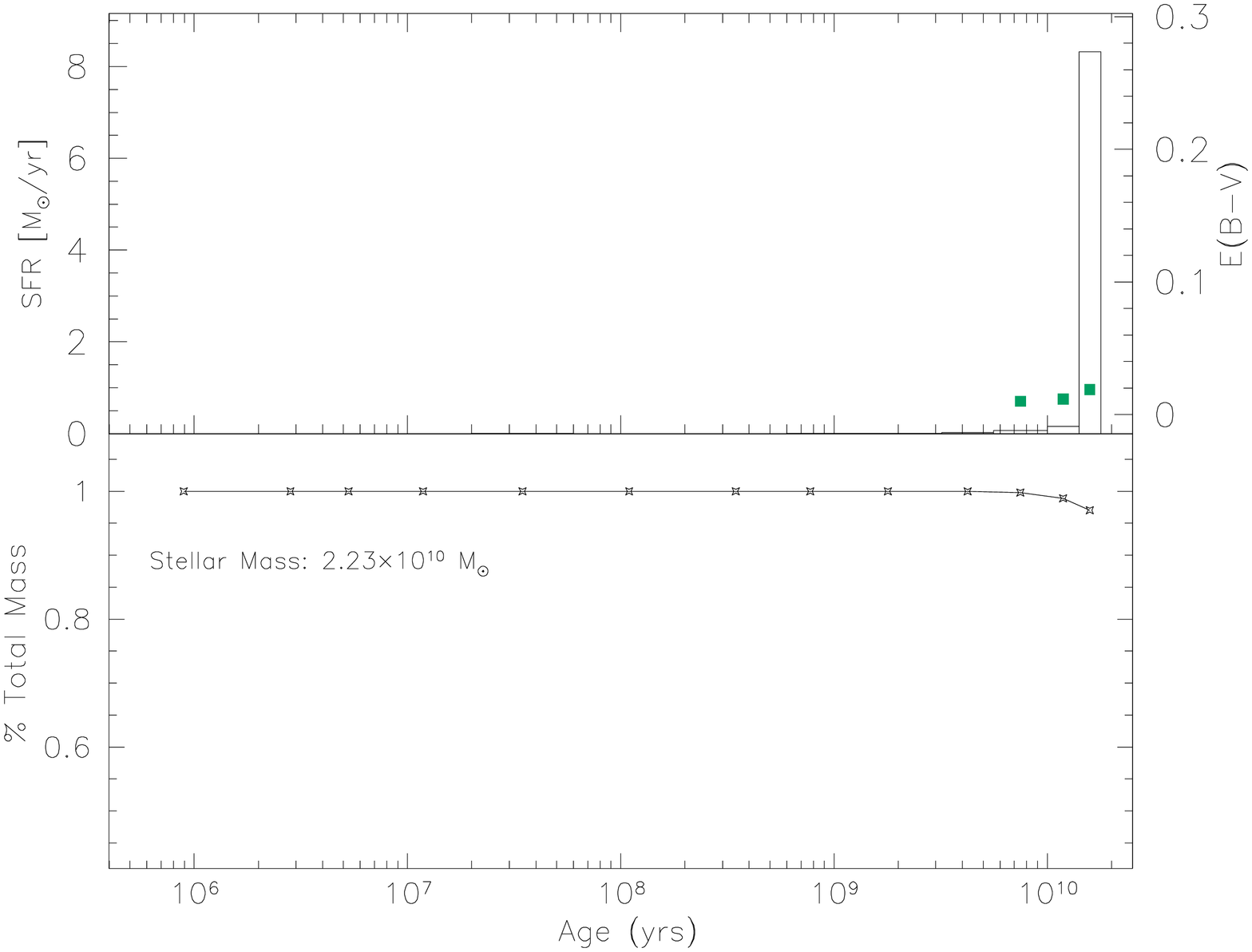}} \\
\end{tabular}
\caption{Model fit to the observed spectrum of a passively-evolving galaxy in the Abell cluster A119. On the left-upper panel the observed spectrum is plotted as a thick, red line, while the model spectrum is shown with a blue, thin line. Spectra are shown in the rest-frame, and normalized to $1$ at $5550$ \AA. Fit residuals are shown in the middle panel, in the same panel, as a black dotted line. Superimposed to this line, the continuum bands used to constrain the fit are shown, together with their error bars. In the left-lower panel we show the spectra of the SSPs that compose the final model, labeled with increasing number as age increases (i.e. $13$ is the oldest). In this case the main contribution to the light and to the stellar mass is given by just one very old stellar population. In the right-upper panel the SFH --i.e. star formation rate in solar masses per year as a function of look-back time-- is shown. The value of SFR is represented by the height of histograms, while their width represents the duration assumed for each population. The extinction value for each SSP is represented by green squares, whose scale is given on the right axis. In the lower right panel the cumulative mass, normalized to the total mass, is shown at various ages. The value of total mass refers, in this case, to the aperture magnitude.}
\label{fig:ell}
\end{figure*}

To illustrate the main outcome of our model, in this section we show two examples of application of the spectral fitting procedure to observed spectra. To this aim, we used two \wings \ spectra of galaxies belonging to the cluster Abell 119 and Abell 1069 respectively. The first one is typical of an elliptical galaxy (Fig. \ref{fig:ell}), with no emission lines, weak hydrogen lines in absorption and deep absorption lines from calcium, sodium and magnesium. Using the observed V-band magnitude measured at the same position and with the same aperture of the fiber used for spectroscopic observations, the mass of stars within the fiber aperture, as derived for this particular solution, is found to be $8.0 \times 10^9$ M$_\odot$ while a value of $1.09 \times 10^{11}$ M$_\odot$ is found using the total V-band magnitude as normalization for the spectrum.

\begin{figure*}
\centering
\begin{tabular}{l l}
\rotatebox{270}{\includegraphics[height=0.51\textwidth]{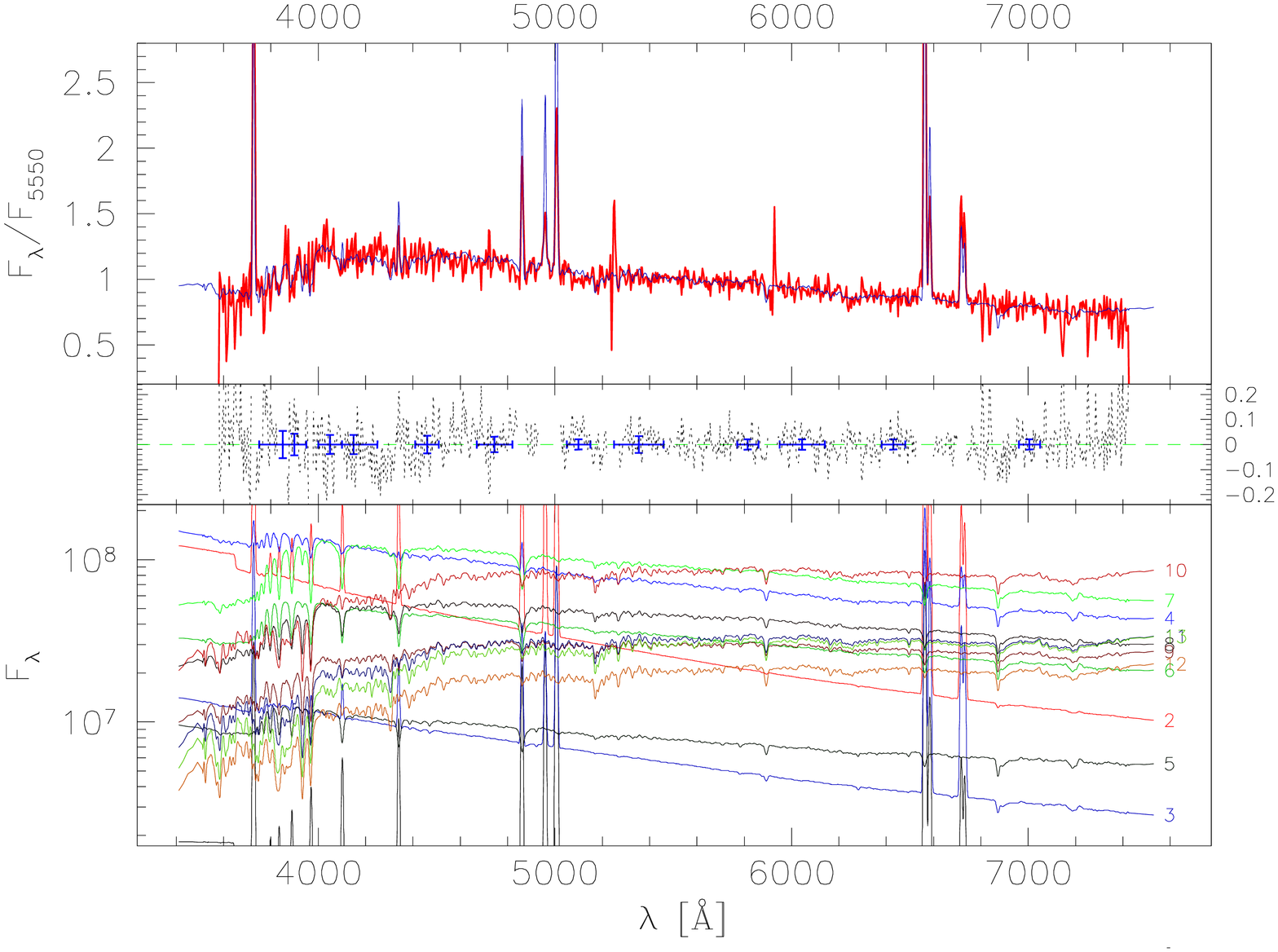}} &
\rotatebox{270}{\includegraphics[height=0.5\textwidth]{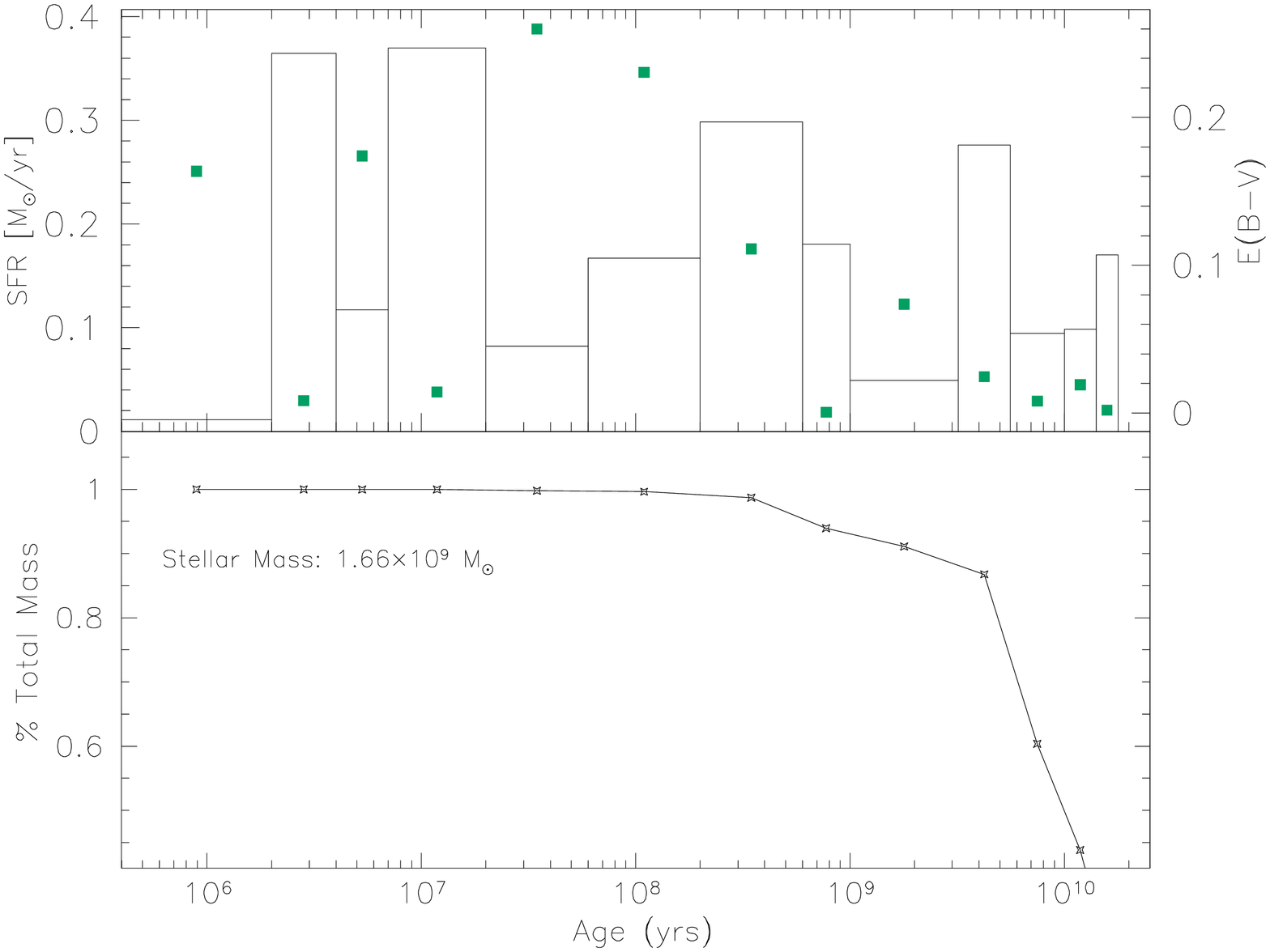}} \\
\end{tabular}
\caption{Fit of an emission-line spectrum of a galaxy in the Abell cluster A1069. Symbols and lines as in Fig. \ref{fig:ell}. Note that residuals --in correspondence of spectral lines that are being fitted-- are not shown, since the comparison model-observed is made on EW values and not on the line profile. Note also that other spectral regions where the residuals are far above the value of the observed error, include non--fitted lines (such as \Oiii \ at $4959$ and $5007$ \AA, respectively), those in correspondence of sky lines, and part of the ultraviolet continuum which, since it is a spectral region which is particularly prone to calibration uncertainties, is not taken into account in the fitting constraints.}
\label{fig:elines}
\end{figure*}

The second spectrum was chosen as an example of a star-forming galaxy, with a blue continuum and strong emission lines (Fig. \ref{fig:elines}). The V-band fiber-aperture magnitude yields a stellar mass value of $1.45 \times 10^9$ M$_\odot$, while the total magnitude gives $8.57 \times 10^9$ M$_\odot$.

\wings \ spectra have been obtained with the 2dF spectrograph at the Anglo-Australian Telescope with a fiber aperture of $2"$ and a spectral resolution of  $9$ \AA \ for the south sample, while observations of the north-sample were performed using the WYFFOS spectrograph at the William Herschel telescope, with a fiber aperture of $1.6"$ and a spectral resolution of $6$ \AA.

The right panels of Figs. \ref{fig:ell} and \ref{fig:elines} show the main outcome of our analysis, i.e. the extinction (green squares) and the SFH (histograms) or, equivalently, the stellar mass formed at each epoch and the build-up of the stellar mass with time. The left panels show the fit (upper panel) and (lower panel) the decomposition, as performed by our model, of the contribution of stars of various ages to the integrated spectrum. Note that the residuals between model and observed spectrum (panel in the middle) are not shown in correspondence to emission lines, since the line widths for the model spectrum is somehow arbitrary while in the observed one it depends on the velocity dispersion of the gas. Residuals in correspondence of the sky background are also omitted. Emission lines like e.g. \Oiii \ and {\sc [Sii]}, are not included in the set of constrained features because they strongly depend by other physical parameters such as metallicity and gas temperature and density, and they are hence not fitted. Furthermore, other spectral regions of the continuum which are affected by telluric bands, are not included in the set of constraints.

Note also that even though the solutions found for these two spectra are very good in terms of reduced $\chi^2$ values, they are not unique. Different combinations of SFRs and extinction could in general yield equally good results, with similar --or even better-- $\chi^2$. In fact, the most critical aspect when attempting to reconstruct the SFH of a galaxy from an integrated optical spectrum, is the degeneracy in the SFH patterns. This arises basically from the characteristics of our problem: the limited spectral coverage of our data (typically from  $\sim 3700$ to $\sim 7000$\AA \ at rest-frame), and the fact that --as the ages of the stellar populations increases-- their main spectral features become more and more similar for spectra with ``contiguous'' ages. The presence of spectral noise, that can actually modify the main features of a spectrum, further complicates our task.

Note that the pattern of SFR (or, equivalently, mass) and extinction derived by the spectral fitting in the two examples showed in Figs. \ref{fig:ell} and \ref{fig:elines}, are not meant to represent the real SFH and extinction of these galaxies, but are shown just as examples of fitting. In the following section we will discuss in detail how to interpret the results and what type of information can be reliably obtained from fits such as those shown.\\ 

\section{TESTING THE MODEL}

The reliability of the technique has been tested by applying it both to synthetic SFH patterns (this approach is similar to the one used by \citealt{cidfernandes05}) and to observed spectra already studied in the literature. To this aim, we performed two different tests. We first applied the model to optical spectra for which the SFH pattern was already known. To this purpose we built a large set of galaxy synthetic spectra with a very varied range of SFHs, which we used as templates.

On the other hand, we tested the robustness of the method by comparing our results with those obtained using different data and/or techniques (e.g. K03 analysis of \sdss \ data) onto a sample of \wings \ galaxies that were in common with the \sdss \ survey.

\subsection{Test on Templates}

In this section we describe the characteristics of the set of template spectra used in our tests and the results achieved. 

We explored several classes of different star formation histories, aiming to reproduce the different types of SFH that can be found in real galaxies. We included a) SFHs mimicking the histories of galaxies of various morphological types along the Hubble sequence from ellipticals to spirals, b) galaxies with ongoing starbursts of different intensities and extinctions, c) a set of post-starburst spectra whose star formation was truncated at different ages, d) ``random'' SFHs with a series of bursts of star formation at different epochs throughout their evolution and e) SFRs constant with time and truncated at increasingly younger ages. A total of $26$ different SFH patterns were used to build an equal number of spectra. The spectra obtained in this way were then smoothed in order to match the spectral resolution of the \wings \ spectra in the southern hemisphere ($FWHM=9$ \AA, which is the worst case) and were cut in wavelength in order to match the \wings \ spectral coverage. Finally we added noise to all the spectra, using the task {\tt mknoise} within {\sc iraf}, to reproduce the noise typically found in the lowest S/N \wings \ spectra, i.e. adopting the \wings \ gain and read-out noise values ($2.79$ e$^-$/{\sc adu} and $5.20$ e$^-$, respectively) and the number of observed counts representative of a typical noisy spectrum of ours.

\subsubsection{The Star Formation History Reconstruction}\label{sec:SFHrec}

\begin{figure}
\includegraphics[height=0.51\textwidth]{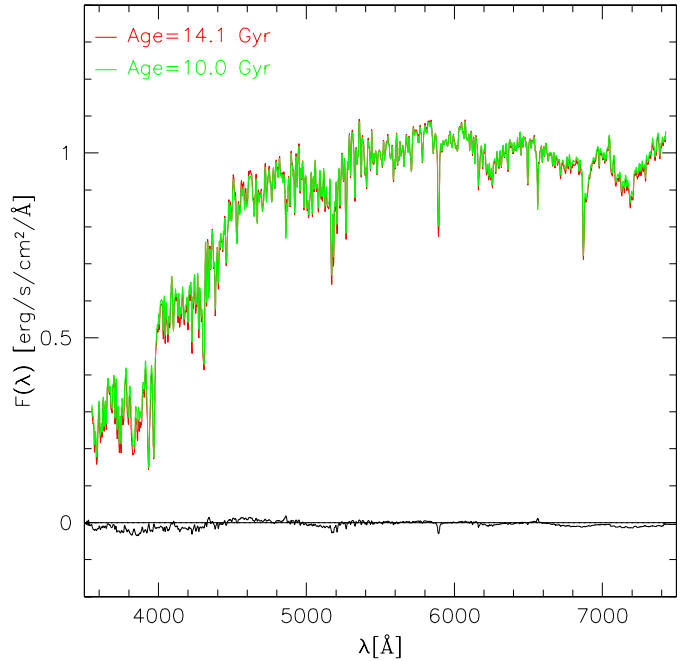}
\caption{Comparison between the spectra of a $14.1$ Gyr (red line) and a $10.0$ Gyr (green line) old SSP. The latter was slightly reddened using the Galaxy extinction law, with an $E(B-V)$ of $0.03$. The two spectra are normalized at $\lambda=5500$ \AA, and their differences are plotted as a black line at the bottom of the plot.}
\label{fig:ssp_comp}
\end{figure}

There is an intrinsic degeneracy in the typical features of spectra of similar age, and this degeneracy increases for older stellar population spectra. In Fig. \ref{fig:ssp_comp} we compare the spectrum of one of the oldest SSPs of our set ($14.1$ Gyr), plotted as a red line, with a slightly younger ($10.0$ Gyr) spectrum from our set, to which a dust attenuation, with an E(B-V) value of $0.03$, was applied. The two spectra are practically indistinguishable in the wavelength range that matches our observations.
There is, hence, an intrinsic limit to the precision of this method in determining the age of the stellar populations that compose a spectrum. Furthermore, the problem worsens if selective extinction patterns are allowed and if data are affected by spectral noise. 

In section \S \ \ref{sec:sspeqw}, we have shown that there are relevant differences in the spectral features as the age changes. While the main features of an observed spectrum cannot be --in general-- adequately reproduced with the use of fewer SSPs, it must be taken into account that a set of observed spectral features can be obtained as a combination of different sets of SSPs parameters (i.e. SFR and extinction). This means that we may be not able, in general, to distinguish between patterns of SFH and extinction that differ in the ages of the dominant stellar populations by small amounts, and also that similar spectral features can be obtained with different linear combination of parameters. We will hence consider as meaningful only the estimates of the stellar mass in four main age bins, that we define as follows:
\begin{itemize}
\item $2\div 20$ Myr: stellar populations characterized by lines in emission and the strongest ultraviolet emission
\item $20\div 600$ Myr: in this age range the Balmer lines reach their maximum intensity in absorption, while the Ca{\sc k} line is almost non detectable
\item $0.6\div 5.6$ Gyr: Balmer absorption lines are, as the age increases, less intense in this age range, while {\sc k} line of calcium reaches its maximum intensity in absorption
\item $5.6\div 17.8$ Gyr: the main features reach an asymptotic value for these SSP
\end{itemize}
We want to stress that this is nothing else but a way of reducing the resolution in terms of ages, of our results. As already pointed out by \cite{cidfernandes07}, the use an over-dimensioned parameter space to fit data, and then its compression to a lower resolution in age, is in fact the expression of a ``principle of maximum ignorance''. The above mentioned characterization of age interval, is the highest resolution that our data can afford, as the analysis on the template spectra has underlined. On the other hand, the adoption of just these four ``SSP'' in a spectral fitting procedure, would imply the assumption that the SFR in the time interval where they are defined, is {\it constant}, which is a far too strong one.

The boundaries of the age bins described above were chosen also based on the results of the SFH reconstruction on the template spectra. These four bins were found to yield the most robust results. Note also that while e.g. \cite{mathis06} are able to define a much finer age grid for their results, considering $6$ bins, we found that for the typical characteristics of our data, $4$ age bins is the maximum resolution we can achieve with a sufficient accuracy.

\begin{figure}
\includegraphics[height=0.525\textwidth]{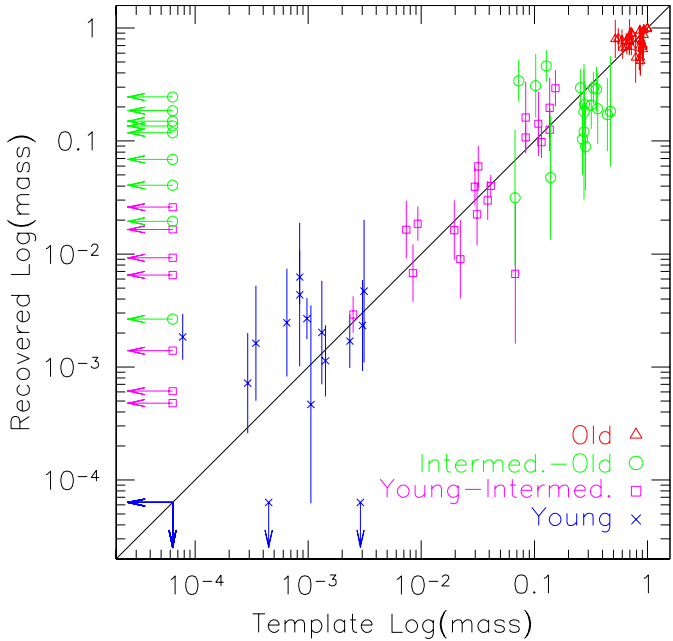}
\caption{Here we compare the values of the mass fraction in the four main age bins, as recovered by means of the spectral fitting, for all of the $26$ template spectra. On the x axis the values of templates are shown, while the values reconstructed by means of the spectral fitting are reported on the y axis. The mass percentage within each age bin are represented with different symbols and colours, together with their error-bars as derived from the fitting technique (see text for detailed explanation). Arrows are used when the mass in a given age bin gives a negligible contribution to the total value. The dashed line represents a perfect agreement between template values and model reconstruction.}
\label{fig:template}
\end{figure}

An example of this approach is shown in Fig. \ref{fig:template} where we compare the SFH of one of our templates to the reconstructed one.

Since the Mass-to-Light ratio of an SSP increases with age, the age bins that are more prone to uncertainties in the mass values and degeneracy effects are the oldest ones. This is due to the fact that changes of a few percentages in the mass of young stars will result in a substantial change of the spectral features, while these changes are more subtle in old stars. There is indeed a non negligible difference in the Mass-to-Light ratio of different old SSPs so that it turns out that the mass recovered by a spectral fitting can strongly depend on the age of the oldest stellar populations included in the fit itself. 

Given the limited spectral range available, we expect a certain degree of degeneracy in the solutions, which are represented by a set of extinction and SFR values. As already pointed out above, this is in part due to the fact that spectra of SSPs close in age are similar. Another source of degeneracy that can arise is caused by extinction. Since we are fully exploiting the selective extinction hypothesis, which allows younger stars to be more dust--reddened with respect to older ones, removing a SSP from the final spectrum can be achieved in two ways: by assigning it a low mass value, or by adopting a high amount of extinction. While the latter picture is likely to happen in Starbursts and objects such as Luminous Infrared Galaxies, this appears to be a non-common situation when applied to the great majority of galaxies in nearby cluster samples which we expect to be mostly passive-evolving objects or in a post-starburst phase. Thus, in the fit of \wings \ spectra we minimized this effect by restricting the maximum values allowed for extinction of the youngest SSPs, to those typical of spiral galaxies, i.e. $E(B-V)\simeq 0.3$, and to increasingly lower values as the age of SSPs increases.

To assess the amount of degeneracy intrinsic to our best fit models and assign an associated error, or uncertainty, to the reconstructed star formation histories, we exploited the features of the adopted optimization algorithm which randomly searches the best combination of parameters that reproduces an observed spectrum (see Sec. \ref{sec:simann}). In fact, starting from a different initial point --in the parameter space-- and using different seeds for the random number generator algorithm, will result in a completely different way of exploring the parameters space itself and could in general lead to a different solution.

As discussed in Section \ref{sec:simann}, the ASA method performs a random exploration of the parameter space, searching for the combination of SFR and extinction values of the various SSPs that minimizes the differences between the model and the observed spectrum. Every move in this space is probabilistically determined but also depends on the moves previously made. 

We performed $30$ SFH reconstruction simulations on each template spectrum. For each one we changed both the initial point and the seeds of the random number generator, in order to evaluate the accuracy and robustness of the results and to check for the presence of spurious trends which could depend on the random nature of the search of a minimum $\chi^2$. 
The final solution depends on the path followed by the algorithm to minimize the $\chi^2$ function, and the path depends in turn on the initial points. Each time we start from a given point we end up in a different position of the parameter space, i.e. on a different solution. Note that all the final configurations have acceptable $\chi^2_\nu$ values. Using the approach described by \cite{pelat98}, we are here looking for solutions to a problem which is underdetermined, i.e. the number of observable constraints is --in general-- less than the number of degree of freedom. Such problems have an infinite number of solutions, and they are all contained within a polyhedron in the parameters space. It would be enough, for us, to find solutions as close as possible to the vertices of the solutions polyhedron, since they would represent extreme values of the quantities we are looking for. Our solution will hence be somewhere within this subspace.

As reference values for the physical quantities of interest --i.e. the total mass, extinction and mass in the $4$ age bins, etc.-- we use those belonging to the model for which the value of total mass is also the median of the distribution of total masses within the $30$ runs. We will refer to this model as the {\it reference model}. A good estimate for the error, to be associated with these quantities, is given by the following:
\begin{equation}
\Delta Q=\frac{Q_{max}-Q_{min}}{2}
\end{equation}
where $Q_{max}$ and $Q_{min}$ are the value of the physical quantity belonging to the model with the highest and lowest stellar mass, respectively, among the $30$ runs. With this estimate of the uncertainties, we are sure to include all the possible acceptable solutions, since we are taking as extreme, those values belonging to models at the vertices of the polyhedron of solutions.

\begin{figure*}
\rotatebox{270}{
\includegraphics[height=1.01\textwidth]{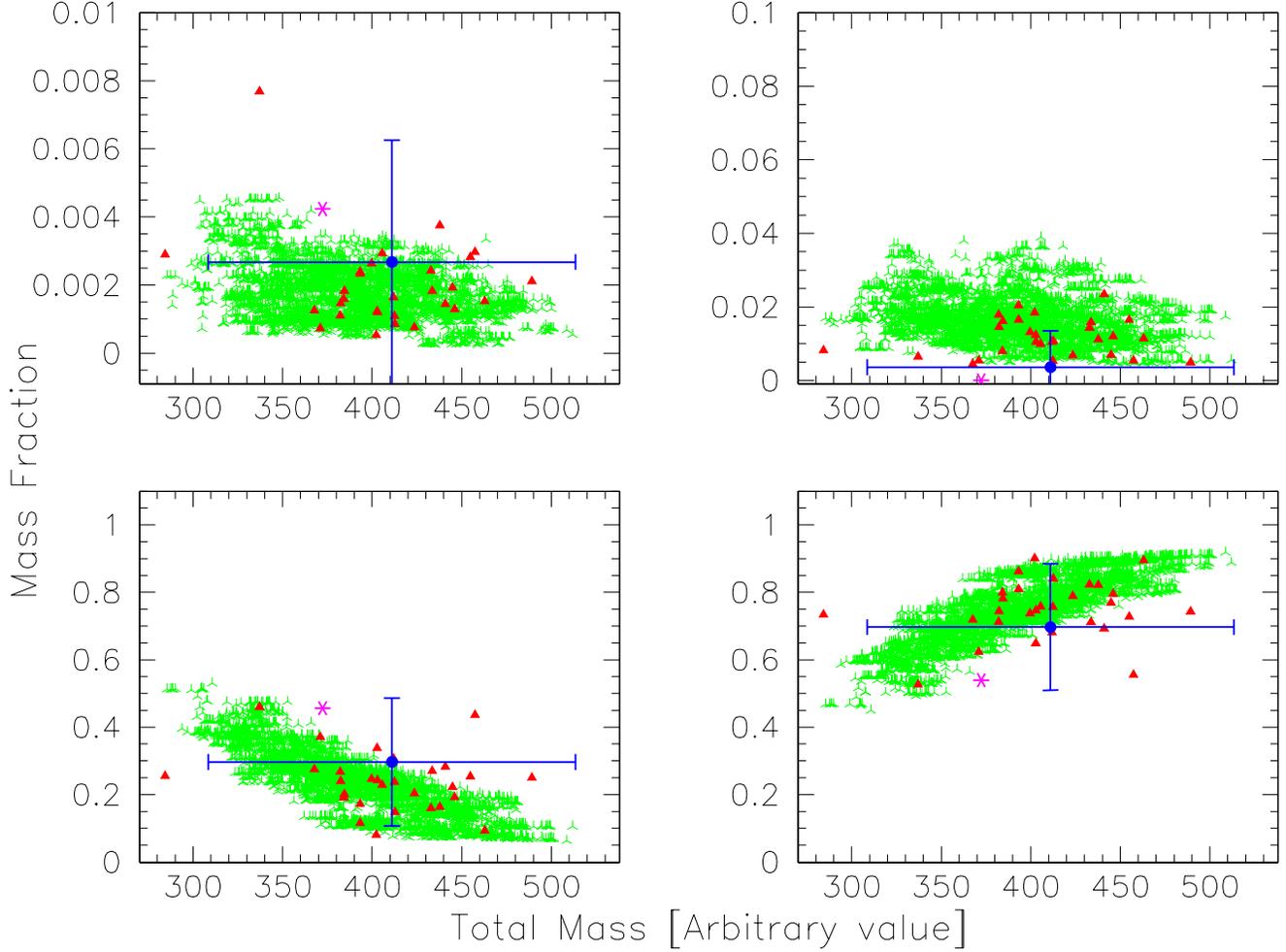}}
\caption{We report in these plots the results of the minimum $\chi^2$ search on the $RND_1$ template (spectrum no. $24$, see also table \ref{tab:templates}), which is actually one of those that shows the largest discrepancy between ``real'' and reconstructed values. On the x axis we plot in all the boxes the total mass, while in the y axis we plot the mass fractions of the youngest to the oldest age bin, respectively from the upper-left to the lower-right panel. The starred point (magenta) represents the values of the template, red triangles refer to the best-fit models over $30$ runs and the blue circle is the reference model, with error bars computed as described in the text. We plot also the mass and mass fractions of those models with $\chi^2_\nu \leq 2\cdot\chi^2_{\nu,min}$ taken from the reference model's best fit search.}
\label{fig:bestchi}
\end{figure*}

As an example, in Fig. \ref{fig:bestchi} we summarize the results of the spectral fitting for the template spectrum no. $24$ (SFH with instantaneous bursts at random ages): the magenta starred point represents the real value of mass and mass fractions, the blue circle represents the position of the ``reference model'' (value and error bars are computed as explained above), red triangles are the values of the best fit model for each one of the $30$ runs, while green points represent all the models for which $\chi^2_\nu \leq 2\cdot\chi^2_{\nu,min}$ (where $\chi^2_\nu$ is the reduced $\chi^2$) in the reference model best fit search.

In table \ref{tab:templates} we report both the mass fractions in the four age bins of the $26$ templates, and the values of mass fractions as reconstructed by the model, over a total number of $30$ runs, together with their uncertainties.
\begin{table*} 
\centering
\begin{tabular}{l|c||c|c|c||c|c|c||c|c|c||c|c|c}
Name  &{\sc id }& \multicolumn{3}{c||}{Bin 1} & \multicolumn{3}{c||}{Bin 2} & \multicolumn{3}{c||}{Bin 3} & \multicolumn{3}{c}{Bin 4}  \\
\cline{3-14}
       &    & Templ.&  Rec. &$\Delta$& Templ.&  Rec.&$\Delta$& Templ.&  Rec.&$\Delta$& Templ.&  Rec.&$\Delta$ \\
\hline
Elliptical   & 1 & 0.000 & 0.000 & 0.000 & 0.000 & 0.001 & 0.001 & 0.000 & 0.007 & 0.026 & 1.000 & 0.992 & 0.026   \\
S0           & 2 & 0.002 & 0.003 & 0.001 & 0.008 & 0.003 & 0.005 & 0.072 & 0.276 & 0.180 & 0.918 & 0.718 & 0.183   \\
Sa           & 3 & 0.002 & 0.001 & 0.001 & 0.002 & 0.003 & 0.001 & 0.067 & 0.111 & 0.095 & 0.929 & 0.885 & 0.095   \\ 
Sb           & 4 & 0.002 & 0.001 & 0.005 & 0.019 & 0.022 & 0.013 & 0.256 & 0.198 & 0.138 & 0.723 & 0.779 & 0.131   \\
Sc           & 5 & 0.002 & 0.003 & 0.003 & 0.031 & 0.045 & 0.020 & 0.318 & 0.296 & 0.129 & 0.649 & 0.656 & 0.132   \\ 
Irr          & 6 & 0.003 & 0.002 & 0.004 & 0.038 & 0.021 & 0.014 & 0.020 & 0.432 & 0.206 & 0.859 & 0.546 & 0.200   \\ 
Cnt. SFR$_1$ & 7 & 0.000 & 0.000 & 0.000 & 0.000 & 0.002 & 0.006 & 0.268 & 0.153 & 0.111 & 0.732 & 0.845 & 0.110   \\ 
Cnt. SFR$_2$ & 8 & 0.000 & 0.000 & 0.000 & 0.000 & 0.008 & 0.005 & 0.285 & 0.016 & 0.130 & 0.715 & 0.976 & 0.130   \\
Cnt. SFR$_3$ & 9 & 0.000 & 0.000 & 0.000 & 0.022 & 0.006 & 0.011 & 0.278 & 0.174 & 0.129 & 0.700 & 0.820 & 0.137   \\
Cnt. SFR$_4$ &10 & 0.000 & 0.000 & 0.000 & 0.030 & 0.033 & 0.018 & 0.276 & 0.094 & 0.361 & 0.694 & 0.873 & 0.357   \\
Cnt. SFR$_5$ &11 & 0.000 & 0.000 & 0.000 & 0.032 & 0.059 & 0.030 & 0.276 & 0.153 & 0.250 & 0.693 & 0.788 & 0.242   \\
Cnt. SFR$_6$ &12 & 0.000 & 0.000 & 0.000 & 0.000 & 0.001 & 0.001 & 0.139 & 0.020 & 0.121 & 0.861 & 0.979 & 0.121   \\
Cnt. SFR$_7$ &13 & 0.000 & 0.000 & 0.000 & 0.000 & 0.000 & 0.000 & 0.000 & 0.027 & 0.030 & 1.000 & 0.973 & 0.030   \\
Flat         &14 & 0.002 & 0.001 & 0.001 & 0.042 & 0.025 & 0.010 & 0.354 & 0.217 & 0.165 & 0.602 & 0.757 & 0.168   \\  
P.SB$_1$     &15 & 0.000 & 0.000 & 0.000 & 0.008 & 0.016 & 0.013 & 0.472 & 0.267 & 0.380 & 0.520 & 0.717 & 0.381   \\
P.SB$_2$     &16 & 0.000 & 0.000 & 0.000 & 0.010 & 0.016 & 0.008 & 0.333 & 0.200 & 0.116 & 0.657 & 0.784 & 0.121   \\ 
P.SB$_3$     &17 & 0.000 & 0.000 & 0.000 & 0.109 & 0.142 & 0.131 & 0.103 & 0.308 & 0.280 & 0.788 & 0.550 & 0.376   \\
P.SB$_4$     &18 & 0.000 & 0.000 & 0.000 & 0.154 & 0.224 & 0.131 & 0.000 & 0.122 & 0.334 & 0.845 & 0.654 & 0.379   \\ 
P.SB$_5$     &19 & 0.002 & 0.000 & 0.000 & 0.084 & 0.107 & 0.037 & 0.000 & 0.186 & 0.233 & 0.914 & 0.707 & 0.236   \\ 
P.SB$_6$     &20 & 0.003 & 0.004 & 0.015 & 0.084 & 0.110 & 0.173 & 0.000 & 0.294 & 0.289 & 0.913 & 0.592 & 0.316   \\
SF$_1$       &21 & 0.001 & 0.004 & 0.004 & 0.115 & 0.085 & 0.038 & 0.000 & 0.081 & 0.197 & 0.884 & 0.830 & 0.213   \\ 
SF$_2$       &22 & 0.002 & 0.007 & 0.014 & 0.136 & 0.169 & 0.166 & 0.000 & 0.052 & 0.186 & 0.862 & 0.772 & 0.241   \\
SF$_3$       &23 & 0.002 & 0.007 & 0.005 & 0.136 & 0.152 & 0.069 & 0.000 & 0.068 & 0.152 & 0.862 & 0.773 & 0.204   \\
RND$_1$      &24 & 0.003 & 0.003 & 0.004 & 0.000 & 0.003 & 0.010 & 0.443 & 0.297 & 0.190 & 0.554 & 0.697 & 0.187   \\
RND$_2$      &25 & 0.002 & 0.002 & 0.001 & 0.000 & 0.028 & 0.023 & 0.127 & 0.380 & 0.176 & 0.871 & 0.590 & 0.176   \\
RND$_3$      &26 & 0.002 & 0.003 & 0.001 & 0.067 & 0.025 & 0.021 & 0.000 & 0.212 & 0.223 & 0.931 & 0.760 & 0.217   \\
\hline
\end{tabular}
\normalsize
\caption{In this table we summarise the SFH of the $26$ templates used in our tests -in the {\it Templ.} column, while in the {\it Rec.} column the value as it was found by means of our model. Values refer to the four main age bins defined in the text (where Bin1 is the youngest and Bin4 the oldest), and they are normalized to the total mass.}
\label{tab:templates}
\end{table*}
As can be easily seen, not all the values of mass fraction are adequately reproduced within the errors. We interpret these discrepancies as systematic error which depends on the spectral type, i.e. on the SFH of the galaxy being analyzed. 
The most robust estimates are given for the total stellar masses, any discrepancies being due mainly to the poor determination of the precise age of the stellar populations that constitute the bulk of the luminous mass. Further uncertainties can arise when a spectrum is dominated by young stars, in which case the light of older stars is dominated by the emission of the lower M/L populations. In these cases the mass value is typically underestimated.

The issue of the stellar mass in age bins as we defined them, is more tricky. The main problem in this case is that while the age bins are defined with abrupt interruptions at given ages, the features of SSP which are contiguous in age, but where assigned to different bins, are subject to only smooth changes. So it is actually possible to adequately reproduce an observed spectrum with stellar populations belonging to different bins.

\begin{figure}[!t]
\rotatebox{270}{
\includegraphics[height=0.51\textwidth]{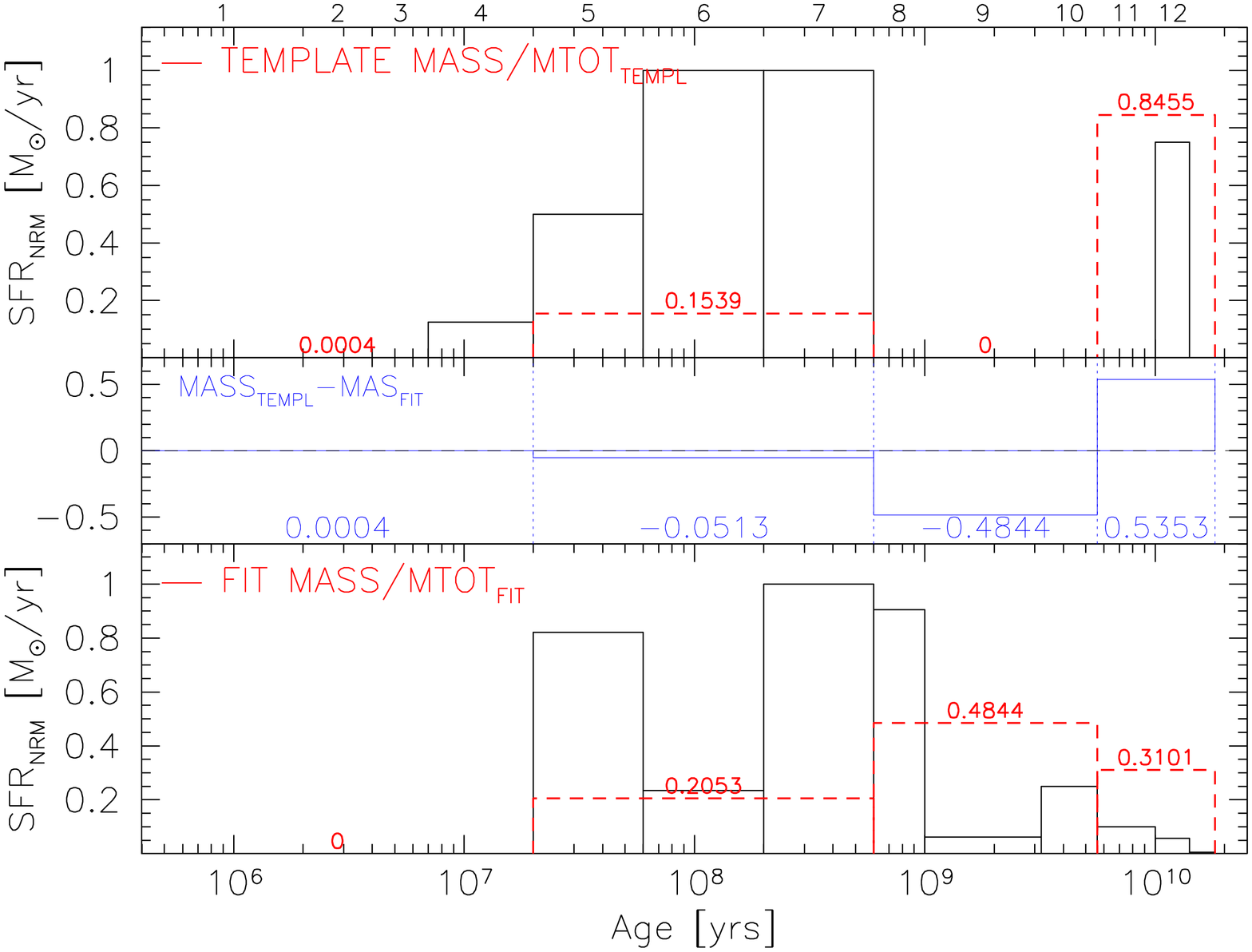}}
\caption{Original and reconstructed star formation pattern for one of our templates. Histograms with a black continuous line represent the intensity of the star formation of a given population of the template, in the top panel, and the one reconstructed by the model in the bottom panel. Histograms in dashed-red line represent the contribution to the total stellar mass of the four age bins --as defined in the text-- for the template and for our reconstruction in the top and bottom panel respectively. In the middle panel the differences in the mass percentage of the four bins are shown. The template used for this example is a post--starburst one (template no. $18$).}
\label{fig:sfhtemp}
\end{figure}
In Fig. \ref{fig:sfhtemp} we show the comparison between the SFH of one of our templates and its reconstruction as performed by our model. This figure depicts the problem of our approach which we just mentioned, i.e. that grouping SSPs of different ages in bins is a reasonable but somewhat arbitrary process. The fact that SSP spectra of contiguous ages are very similar makes them indistinguishable, especially in a spectrum which is given by the composition of populations of many different ages. So, for example in the case showed in Fig. \ref{fig:sfhtemp}, while in the original spectrum there is no contribution from stars in the third age bin, a substantial amount of stellar mass ($\sim 50$\%) is found in this bin after our analysis. The bulk of this mass is anyway found in SSP which are contiguous to the older and the younger bin respectively. Fig. \ref{fig:template} shows the ability of the model in recovering the mass fractions in the $4$ age bins for the $26$ templates.

Note that this way of treating the problem also allows us to take into account the degeneracy of the solutions, since the space of best fit models is explored very well. Actually, we found that a good representation of this space is achieved by limiting ourselves to $\sim 10$ runs. Increasing the number of trials does not increase the accuracy of the exploration in the best models parameters space. So, in practice, the results will be obtained over $11$ runs for each value of the SSP metallicity that we will consider (see \S \ \ref{sec:met}).

We wish to stress that, other than the uncertainties typical of this modeling and those deriving from calibration issues, the results provided by such models in general depend on various other parameters. The most important ones are probably the knowledge of stellar evolution and its treatment from a theoretical point of view and the IMF issue which can significantly alter the values of stellar masses and of mass-to-light ratios. 

\subsubsection{The Influence of Spectral Noise}

As one expects spectral noise to influence the values of measured quantities (e.g. flux on the continuum, EW, etc.), thus affecting the results of the SFH reconstruction, we tackled this issue by applying our method to the original template set, i.e. without the artificial addition of spectral noise. In this way we can roughly test the dependence of our results on the S/N, having the two extreme cases: poor and very high S/N. For the sake of brevity we will only summarize here the main results.

Although the values we recover for the mass percentage in the age bins are in general more precise, and the uncertainties are slightly smaller --in comparison to the results shown in table \ref{tab:templates}-- it is clear that spectral noise is not what drives the degeneracy of solutions and hence the errors we estimate from our analysis: the uncertainties on the mass fraction in the age bins are significantly smaller for the high-quality spectra only for the second age bin. Once again we find that the characteristic features of a spectrum can be reproduced by different mixing of stellar population of different ages: the job can be well done by adjusting the weights, in the final spectrum, of the various SSP.

\subsubsection{The Metallicity Issue}\label{sec:met}

One of the physical parameters that  are commonly derived from spectral synthesis is the metallicity, at least that of the stellar populations that dominates the emission in the spectral domain under investigation. \cite{cidfernandes05} claim the ability of determining metallicities with the actually remarkably good precision of $0.1$ dex. In order to verify the possibilities of our method to recover such information, we performed a SFH reconstruction test on our template set, adopting SSP spectra of 2 other different metallicity values with respect to the original one (we recall that the templates were  built with SSP all of solar metallicity), namely $Z=0.004$ and $Z=0.05$. As for the previous section, here we will only briefly illustrate our results. 

The SFH reconstructions were performed both on high and low S/N template spectra for the $3$ values of the metallicity. A best fit was chosen among those obtained for the various metallicities, as the one which yielded the best $\chi_\nu^2$. We considered not acceptable those fits for which $\chi_\nu ^2 > 3 \cdot\chi_{\nu,best}^2$. 

We found that in high S/N spectra the value of metallicity was correctly recovered in the $\sim 50$\% of the cases, in particular in those templates where the emission is dominated by old stars (such as the case of the elliptical or S0 template and those ones with a small fraction of the total mass coming from stars younger than $10^{10}$ years), while in the other cases at least two metallicity values were found to yield an equally good fit. The issue changes dramatically when the S/N worsens: we could only recover the correct value of metallicity only in $2$ out of $26$ spectra, while the other $2$ metallicities yielded acceptable fits for the rest of spectra. This is easily explained considering the well known age--metallicity degeneracy issue: metal-rich, young systems show features very similar to metal-poor old ones. Playing a bit with the ages of the stellar populations can often be enough to fit a spectrum even with the ``wrong'' metallicity.

Hence, the fact that a spectrum can be actually reproduced by means of SSP with an homogeneous value of the metallicity, but which can actually differ from the ``real'' one, may introduce a further component in the uncertainties on the total mass and on the mass percentage in the age bins. Nevertheless, this uncertainty is already well included in the error bars we estimate with our method, since it automatically explores the possibility of fitting a spectrum choosing among SSP with different ages and amount of extinction mimicking in some way, a metallicity effect.

As a result of this analysis and considerations, we will take, as the reference model for a spectrum, the one with the lowest $\chi_\nu^2$ among those obtained with the $3$ values of metallicity specifying if also other values of the metallicity (between $0.004$ and $0.05$) are equally probable.

\subsection{A Comparison With SDSS Results}

To have a direct comparison with other results in the literature, we performed an analysis on a small sample of galactic spectra taken from the Sloan Digital Sky Survey \sdss. This control sample was chosen among those spectra in common between the SDSS and the WINGS survey in the cluster Abell 119. In this way, we can compare the values of stellar masses, extinction, star formation rate for a total of $23$ spectra.

Given the results in terms of stellar masses and extinction from K03, we performed different kind of comparisons. The model was applied both on \wings \ and \sdss \ spectra taken from the 4th data release. In this way we were able to compare both our data and our method with the \sdss \ survey and to the K03 work, respectively. The method is the one that was described above: a total number of $33$ models --$11$ for each value of the SSP metallicity-- were obtained for each of the $23$ spectra analyzed, and final values of masses, extinctions and errors were taken as explained in section \S \ref{sec:SFHrec}

Following the method adopted by K03, the values of total mass are computed by rescaling the masses obtained by fitting the optical spectra --which had been calibrated on the V-band fiber magnitude for the \wings \ spectra and to g-band petrosian magnitude for \sdss \--, to the total V and g magnitude respectively. In this it is implicitly assumed that there are no spatial gradients of the stellar populations along the galaxies' profiles. 
\begin{figure*}
\centering
\begin{tabular}{c c}
\includegraphics[height=0.50\textwidth]{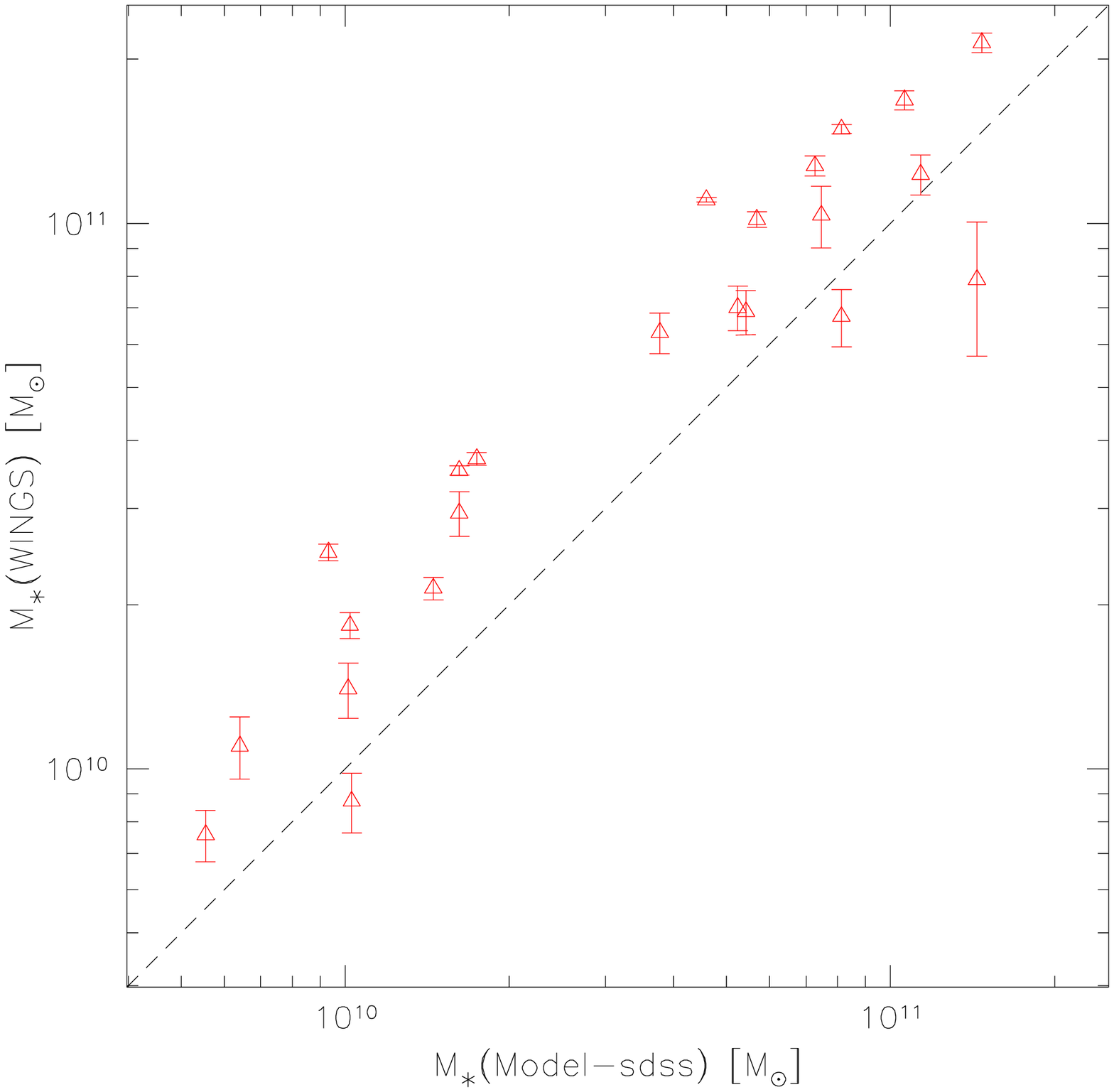} &
\includegraphics[height=0.50\textwidth]{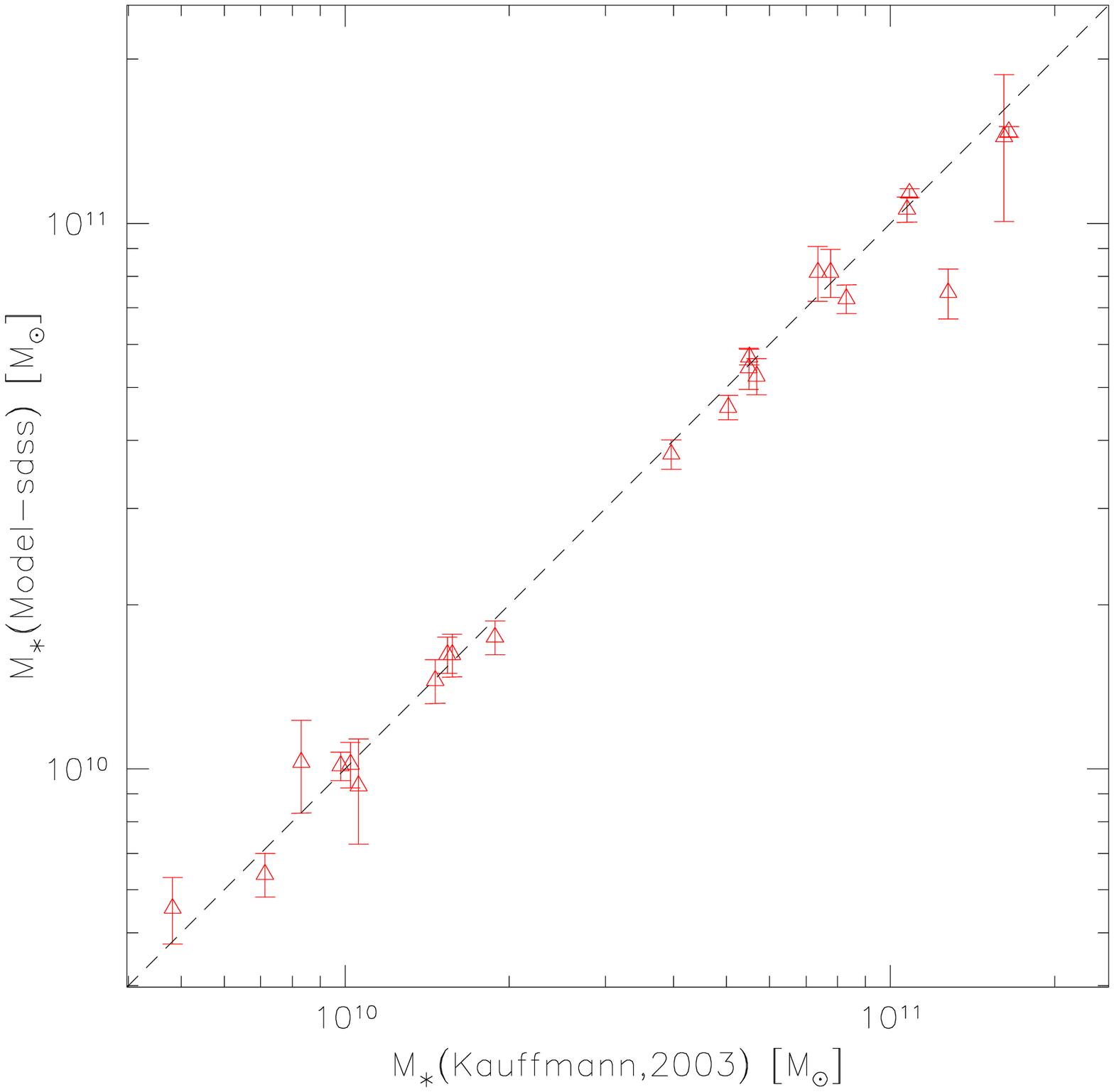} \\
\end{tabular}
\caption{On the left-panel plot we report the comparison between the values of total stellar masses obtained by applying our model to a subsample of spectra which are in common between the \wings \ and the \sdss \ surveys. The error-bars relative to the determination of mass in our model, are computed as explained in section \ref{sec:SFHrec}. Although there is a clear trend in the values, there is an evident off-set, due to the fact that \wings \ magnitudes are typically brighter than the \sdss \ ones. On the right panel a similar comparison is performed between mass values as obtained by K03 work, and our model's values which were obtained using \sdss \ data. In spite of differences in the analysis methods, \sdss \ data yield fully compatible results.}
\label{fig:wings-sdss}
\end{figure*}
In Fig. \ref{fig:wings-sdss} we compare and summarize the results of our analysis. On the left panel we show the values of total stellar masses as computed by our model applied both on \wings \ and \sdss \ data. The constant off-set in the mass values is due to the fact that \wings \ magnitudes are typically brighter than \sdss \ ones. On the right panel we show the remarkably very good agreement between the values yielded by the application of our model to \sdss \ data and the results from the work of K03 (the values here reported are {\it dust-corrected} mass values). The spectra in this subsample are all dominated by old stars, so that a comparison of the mass formed in bursts younger than $10^9$ years could not be performed. Note that our method is generally able to obtain a higher level of detail when reconstructing the SFH.

Note that while our set of SSPs was built with a standard Salpeter IMF (masses in the range $0.15 \div 120$ M$_\odot$), K03 use a Kroupa IMF: this would imply a difference by a factor of $\sim 1.3$ in the masses. So the values reported in Fig. \ref{fig:wings-sdss} have been corrected to account for this difference. 

\section{SUMMARY AND CONCLUSION}

We have modified, improved and extended an already existing spectrophotometric code, which is able to reproduce the main features of an observed galactic spectrum and that includes also emission lines from diffuse gas. Particular care was taken when measuring the EW of lines in the observed spectra, since they are of fundamental importance when trying to determine the features of stellar populations. A new method was developed to automatically measure the most important spectral lines --both in emission and in absorption--, avoiding the use of fixed pass-bands which yields values that can be prone to large uncertainties.

We performed accurate tests of our model on templates spectra, which were built on-purpose spanning a whole range of SFH, in order to test the reliability of our SFR reconstruction. While a fine age subdivision was used to reproduce the spectra, it is unfeasible to try to recover the stellar mass values in such a detail. This is in part due to the presence of spectral noise, which alters the features that characterize the stellar populations of various ages, and in part to the fact that the elder the stars the smoother the differences between one another. 

Two different kind of tests where performed: 1) analysis on template spectra --built in a way to resemble the characteristics and the quality of \wings \ spectra-- for which the SFH pattern was already known; 2) analysis on observed, already studied spectra. The first test phase was done both on low and high S/N spectra, in order to verify the dependence, if any, of the quality of results from the spectral noise. This tests showed that spectral noise seems to introduce second-order uncertainties, the main factor being an intrinsic degeneracy due both to the similarity of old SSP spectra and to the limited spectral range at disposal for our analysis. Furthermore we tested the ability of the code in determining the metallicity. This is actually a task which is heavily dependent on spectral noise: the model was able to recover the correct metallicity in $\sim 12$ out of $26$ spectra in high quality spectra, and only in $2$ for the low quality ones. In other cases it is evident that any difference in metallicity can be compensated with an adjustment in the stellar population age.

Note that, in principle, when different values of the metallicity are found to provide equally good fits to an observed spectrum --so that nothing can be said on this issue--, the uncertainties on both masses (total and in the age bins) and extinction can increase. We verified that in any case the error bars provided by our method for these physical parameters, reasonably take into account for this further source of uncertainty.

The second test-phase on observed spectra in common with the \sdss \ project, helped to verify the reliability of the model in absolute terms, by comparison with the results obtained by other works. The comparison with results obtained by K03 was found to be remarkably good. 

This analysis will provide star formation histories obtained from the model fits and a catalog of measured EW for the most important spectral lines, automatically computed with a new, highly reliable method, for all galaxies with spectra from the \wings \ survey.



\begin{appendix}
\section{The ASA Optimisation Algorithm}

Starting from a given point in this $S$--dimensional space we want to find a point which is as close as possible to the absolute minimum, trying to avoid local minima. The ``philosophy'' of the Adaptive Simulated Annealing method resides in a thermodynamic analogy that is the way in which metals re-organize their internal structure when they are cooled. When the molecules of a given material have a certain temperature, they are free to move away one from the other, but if the material is cooled slowly enough this mobility is lost, and atoms reach a pure crystal configuration. This represents a state of minimum energy for the system. If such systems are cooled very slowly they can spontaneously reach this status. Thus we can consider the function we want to optimize (which in this case is given by equation \ref{eqn:mf}) just like the energy of our system. 

The Boltzmann probability distribution, given by:
\begin{equation}
P_E \sim e^{-\frac{E}{KT}}
\end{equation}
expresses the concept that even when a thermodynamic system is found to be in thermal equilibrium at a temperature {\bf T}, there is a non--zero probability to find the system itself in a high energy configuration (with respect to the energy defined by $KT$). Consequently, there is the corresponding possibility that the system moves from the local minimum in the energy value to this higher energy configuration to subsequently reach a state with lower energy. In other words the system can both rise and lower its energy, but the lower the temperature, the less likely the system will move towards higher energy states. As already pointed out, in our case the energy of the system is the function that is being minimized while the temperature represents the boundaries in which our parameters --values of extinction and mass-- are relegated, and these parameters can be visualized as the physical configuration of atoms within the crystal.

Following this analogy the optimization process begins from a given starting point --i.e. in a given value of the function that is being optimised -- performing then a number of changes in its configuration, moving from energy $E_1$ to $E_2$ with a probability which is defined by:
\begin{equation}\label{eqn:boltz}
P = e^{-\frac{E_2-E_1}{KT}}
\end{equation}
Note that if $E_1>E_2$, $P$ is larger than $1$ and in this case the probability of a change is set equal to $1$. That is: the system always chooses to move towards those configurations corresponding to lower energy values, but there is also a non--zero probability that the system moves from the last minimum to other points where energy is higher. From the previous equation we see how the temperature becomes a deciding factor for movements choice. As $T\rightarrow 0$ the probability for the code to choose uphill moves will tend to $0$, and the system will rest on the last point of minimum found. Starting from a given point, the ASA algorithm randomly chooses a ``trial point'' lying within the range of the permitted values defined by a range vector which is used to restrict the movements in the parameter space. The range of variation is not kept constant but is allowed to be resized at each step, and its minimum and maximum values are defined by the user. The function in Eq. \ref{eqn:mf} is evaluated at this point and its value is compared to that corresponding to the previous point. Downhill moves are always accepted and the algorithm continues starting from that trial point. Uphill moves may also be accepted; the decision is made by the so called ``Metropolis criterion'' that uses the ``temperature'' and the size of the uphill move in a probabilistic manner. The smaller the temperature and  the size of the uphill movements are, the more likely that movement will be accepted. If the trial is accepted, the algorithm moves on from that point. If it is rejected, another point is chosen instead for a trial evaluation. A fall in the temperature is then performed upon the system, governed by a temperature reduction factor $R_T$, according to the following:
\begin{equation}
T(i+1) = R_T \cdot T(i) 
\end{equation}
where $i$ represents the iteration index. The value of the temperature reduction factor is always set to be $0.85$. As the temperature decreases, uphill moves are less likely to be accepted causing the percentage of rejected moves to grow. So it happens that as the temperature declines, the variation range of parameters is restricted and the Simulated Annealing algorithm focuses upon what it turns to be the most promising area for optimization.

The power of this method also resides in the {\it re--annealing} option which can account for the changes in sensitivity: when a new deep minimum is found, the ranges of variation of parameters are increased to avoid trapping the system in this deep minimum.

The termination criterion is achieved when the values of the merit function --i.e. our $\chi^2$--, during the last $6$ temperature variations, differ from the one at the current temperature by less than a user-defined value.
\end{appendix}

\end{document}